\documentclass{article}

\usepackage{arxiv}
\usepackage[numbers]{natbib}
\usepackage[utf8]{inputenc} % allow utf-8 input
\usepackage[T1]{fontenc}    % use 8-bit T1 fonts
\usepackage{hyperref}       % hyperlinks
\usepackage{url}            % simple URL typesetting
\usepackage{booktabs}       % professional-quality tables
\usepackage{amsfonts}       % blackboard math symbols
\usepackage{nicefrac}       % compact symbols for 1/2, etc.
\usepackage{microtype}      % microtypography
\usepackage{lipsum}		% Can be removed after putting your text content
\usepackage{graphicx}
\usepackage{natbib}
\usepackage{doi}
\usepackage{amsmath}
\usepackage{diagbox}
\usepackage{longtable}
\usepackage{hhline}
\usepackage{multirow}
\usepackage{rotating}
\usepackage{colortbl}
\usepackage{todonotes}
\usepackage{graphicx}
\usepackage{subcaption}
\usepackage{algorithm}
\usepackage{algpseudocode}
\usepackage{tikz}
\usetikzlibrary{shapes.geometric, arrows, positioning, fit}

\title{MAC-Gaze: Motion-Aware Continual Calibration for Mobile Gaze Tracking}

\date{May 28, 2025} 					% Or removing it

\author{
\href{https://orcid.org/0000-0002-0697-7942}{\includegraphics[scale=0.06]{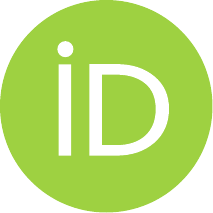}\hspace{1mm}Yaxiong Lei} \\
University of St Andrews \\
St Andrews, Fife, UK, KY16 9SX \\
\texttt{yl212@st-andrews.ac.uk} \\
\And
\href{https://orcid.org/0009-0006-3095-8385}{\includegraphics[scale=0.06]{orcid.pdf}\hspace{1mm}Mingyue Zhao} \\
University of St Andrews \\
St Andrews, UK, KY16 9SX \\
\texttt{mz201@st-andrews.ac.uk} \\
\And
\href{https://orcid.org/0000-0003-3335-8706}{\includegraphics[scale=0.06]{orcid.pdf}\hspace{1mm}Yuheng Wang} \\
University of St Andrews \\
St Andrews, UK, KY16 9SX \\
\texttt{yw99@st-andrews.ac.uk} \\
\And
\href{https://orcid.org/0000-0003-3697-0706}{\includegraphics[scale=0.06]{orcid.pdf}\hspace{1mm}Shijing He} \\
King's College London \\
London, UK \\
\texttt{shijing.he@kcl.ac.uk} \\
\And
\href{https://orcid.org/0000-0003-4206-710X}{\includegraphics[scale=0.06]{orcid.pdf}\hspace{1mm}Yusuke Sugano} \\
University of Tokyo \\
Tokyo, Japan \\
\texttt{sugano@iis.u-tokyo.ac.jp} \\
\And
\href{https://orcid.org/0000-0001-7051-5200}{\includegraphics[scale=0.06]{orcid.pdf}\hspace{1mm}Mohamed Khamis} \\
University of Glasgow \\
Glasgow, UK \\
\texttt{mohamed.khamis@glasgow.ac.uk} \\
\And
\href{https://orcid.org/0000-0002-2838-6836}{\includegraphics[scale=0.06]{orcid.pdf}\hspace{1mm}Juan Ye} \\
University of St Andrews \\
St Andrews, Fife, UK, KY16 9SX \\
\texttt{jy31@st-andrews.ac.uk} \\
}

\renewcommand{\shorttitle}{Motion-Aware Continual Calibration for Mobile Gaze Tracking}

\hypersetup{
pdftitle={A template for the arxiv style},
pdfsubject={q-bio.NC, q-bio.QM},
pdfauthor={David S.~Hippocampus, Elias D.~Striatum},
pdfkeywords={First keyword, Second keyword, More},
}

\begin{document}
\maketitle

\begin{abstract}
Mobile gaze tracking faces a fundamental challenge: maintaining accuracy as users naturally change their postures and device orientations. Traditional calibration approaches, like one-off, fail to adapt to these dynamic conditions, leading to degraded performance over time. We present \textit{MAC-Gaze}, a Motion-Aware continual Calibration approach that leverages smartphone Inertial measurement unit (IMU) sensors and continual learning techniques to automatically detect changes in user motion states and update the gaze tracking model accordingly. Our system integrates a pre-trained visual gaze estimator and an IMU-based activity recognition model with a clustering-based hybrid decision-making mechanism that triggers recalibration when motion patterns deviate significantly from previously encountered states. To enable accumulative learning of new motion conditions while mitigating catastrophic forgetting, we employ replay-based continual learning, allowing the model to maintain performance across previously encountered motion conditions. We evaluate our system through extensive experiments on the publicly available RGBDGaze dataset and our own 10-hour multimodal MotionGaze dataset (481K+ images, 800K+ IMU readings), encompassing a wide range of postures under various motion conditions including sitting, standing, lying, and walking. Results demonstrate that our method reduces gaze estimation error by 19.9\% on RGBDGaze (from 1.73 cm to 1.41 cm) and by 31.7\% on MotionGaze (from 2.81 cm to 1.92 cm) compared to traditional calibration approaches. Our framework provides a robust solution for maintaining gaze estimation accuracy in mobile scenarios.
\end{abstract}

% keywords can be removed

\section{Introduction}
Real-time gaze tracking and interaction on mobile devices is attracting increasing attention from academics and industry, which is to infer point of gaze (PoG) to the screen from facial images captured by the device's front camera. This is often referred to as \textit{appearance-based gaze tracking}~\cite{cheng2024benchmark}, or \textit{mobile gaze tracking}. This research topic is driven by the high-resolution (HR) cameras and powerful computing processors in smartphones. The HR cameras allow for capturing detailed facial and eye images, leading to improved precision of gaze estimation. The computing process makes it possible to perform real-time data processing and run deep learning models, contributing to a responsive and seamless user experience when interacting with gaze. These technological advancements make it possible for real-time gaze interaction, which is inspiring many applications~\cite{lei2023end} to facilitate user interactions with devices.

\begin{figure}[!htbp]
    \centering
    \includegraphics[width=0.55\linewidth]{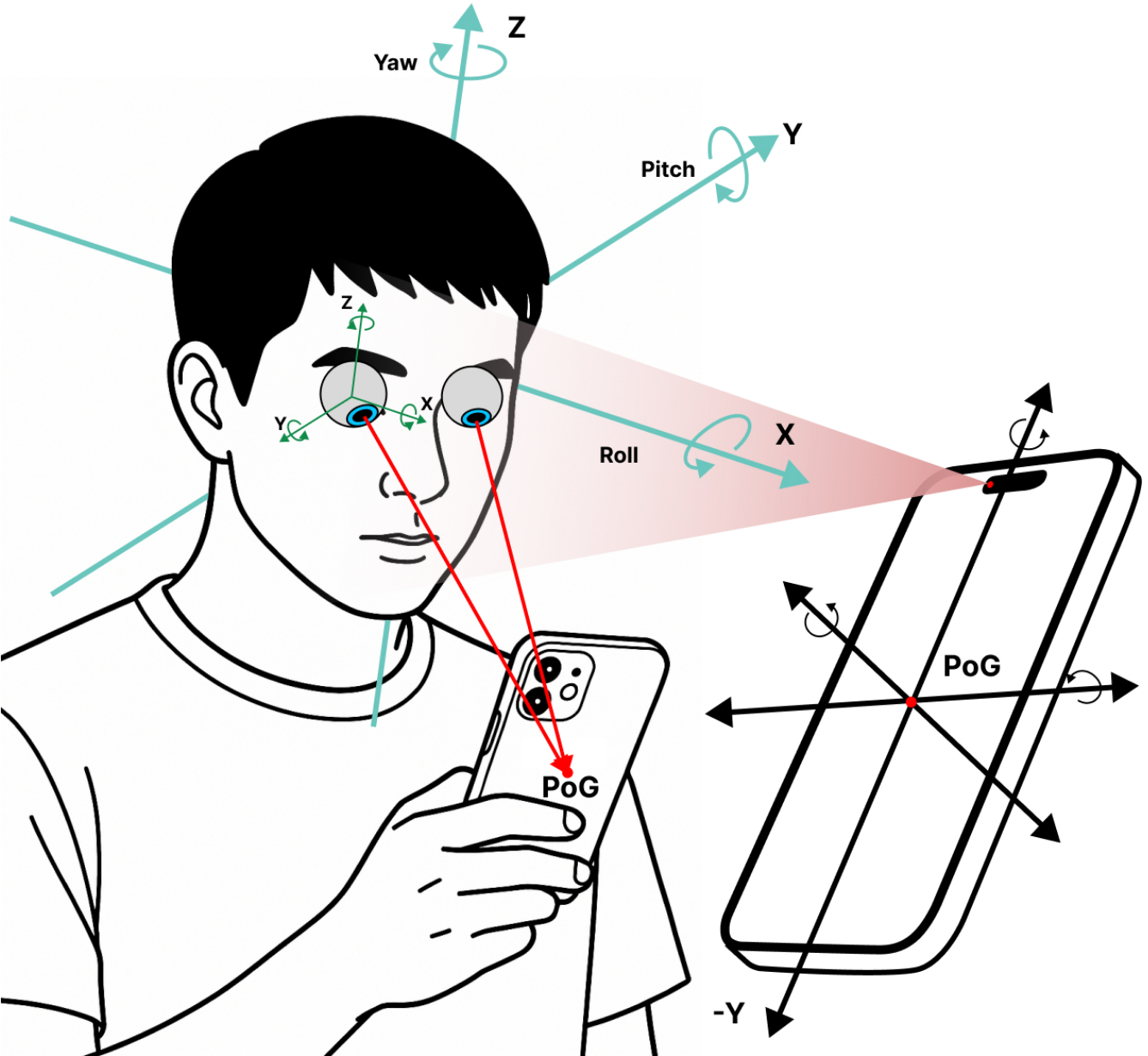}
    \caption{Head-eye-camera configurations in handheld systems that consist of up to 15-DoF from head, eye, and device movements.}
    \label{fig:headeyedof}
\end{figure}

One key element in appearance-based gaze tracking is to learn inherent mapping between facial images and gaze directions, which is regulated by the relative spatial relationships between head, eyes and camera as depicted in Figure~\ref{fig:headeyedof}. 
Traditionally, gaze tracking systems adopt a classic one-off calibration approach; that is, for each user, at the beginning of the use, the system will initiate a calibration process to collect ground truth gaze data; for example, using smooth pursuit to guide users' gaze on dots moving around the boundary of a screen~\cite{lei2023DynamicRead}. With the collected data, a light-weight machine learning model will be trained as a calibrator in order to adjust gaze prediction for the target users~\cite{valliappan2020accelerating,krafka2016eye}. Once calibrated, the system will continue with the same calibrator for future use. If such mapping keeps stable, the gaze tracking can achieve a good level of precision~\cite{lei2023end,cai2025gazeswipe,Huynh2022imon}.

However, this practice does not suit mobile gaze tracking, due to the high DoF and motion-caused input uncertainty in the relative spatial relationship of head, eye, and camera~\cite{zhong2024uncertainty}. In mobile gaze tracking, the head and device each has 6-DoF (degrees of freedom), involving 3 translation (x, y, z) and 3 rotational (pitch, yaw, roll) movements, and the eyes have 3 rotational movements. The degrees of freedom are much higher than AR/VR/glasses gaze tracking where the relative spatial relationship between head and device is fixed and desktop-based gaze tracking where the camera position is fixed. The high DoF makes appearance-based gaze tracking susceptible to changes in user's motion state, their head movements, the way they hold their devices, and the device to head moving speed~\cite{lei2025quantifying}. All these factors lead to degraded accuracy over time~\cite{kong2021eyemu, lei2023DynamicRead, khamis2018understanding}, which is the main obstacle for wider adoption of mobile gaze tracking.

To directly tackle this challenge, we present \textit{MAC-Gaze}, a Motion-Aware continual Calibration approach and address two key questions underlying mobile gaze calibration: \textit{when} to trigger calibration; that is, detecting the \textit{moment} when the recalibration is needed to maintain the high precision of gaze tracking; and \textit{how} to recalibrate; that is, not just for the current motion condition, but incrementally adapting to a diverse range of conditions with the goal of eventually reducing recalibration frequency.

For the \textit{when} question, we leverage the IMU sensors in modern mobile devices. These sensors including accelerometers, gyroscopes, and magnetometers provide rich information about device motion and orientation. Previous studies have utilized IMU data to enhance activity recognition and context-aware applications on mobile devices~\cite{gu2021survey,cao2018gchar}. However, the integration of IMU sensors to trigger automatic recalibration in mobile gaze tracking remains underexplored. In MAC-Gaze we leverage IMU signals to automatically identify when the user's motion state changes (e.g., transitioning from sitting to standing) and monitor fine-grained motion pattern changes (such as subtle variations in device holding posture) that may impact gaze estimation accuracy. We develop a hybrid detection mechanism that combines supervised classification with unsupervised clustering to trigger recalibration accordingly.

For the question on \textit{how} to update, we employ continual learning techniques with a memory buffer to store calibrated data for future update, which helps to reduce catastrophic forgetting of previously encountered motion conditions. Our approach aims to not only improve gaze prediction accuracy but also enhance the user experience by reducing the need for frequent manual recalibrations.

Our work addresses a fundamental challenge in mobile gaze tracking; i.e., the dynamic user-device interaction 
% head-eye-camera relationship 
caused calibration distortion, through an innovative combination of motion sensing, hybrid activity recognition, and continual learning. The resulting system advances the state of the art in mobile gaze estimation, bringing robust gaze tracking closer to widespread adoption in everyday mobile interactions. Our main contributions are listed below:

\begin{enumerate}
    \item We introduce the first multi-modal continual calibration approach (motion and vision) tailored for mobile gaze tracking. By leveraging real-time motion data, our MAC-Gaze method can autonomously identify when recalibration is needed.
    \item We implement a replay-based continual learning mechanism that enables incremental adaptation to evolving motion conditions while preventing catastrophic forgetting of previously encountered states.
    \item We validate our approach through extensive experiments covering diverse motion conditions. Our method demonstrates significant improvements over traditional one-off calibration, reducing error by 19.9\% on RGBDGaze (from 1.73 cm to 1.41 cm) and 31.7\% on MotionGaze (from 2.81 cm to 1.92 cm).
\end{enumerate}

\section{Related Work}\label{sec:related}
In this section, we have reviewed the relevant techniques in gaze estimation with a particular focus on appearance-based gaze estimation, calibration, and human activity recognition.  

\subsection{Gaze Estimation across Devices}
The key impact factor of gaze estimation performance is the spatial relationship between the eye, cameras and target plane. This relationship varies significantly depending on device context, creating distinct challenges and constraining the degrees of freedom (DoF) within the head-eye-camera system.

\textit{Head-mounted systems} (e.g., VR/AR/MR headsets, eye-tracking glasses) offer the most stable conditions, with eye-tracking cameras rigidly fixed relative to the eyes, yielding a 3-DoF scenario limited primarily to eye rotations~\cite{cognolato2018head, tonsen2017invisibleeye}. This configuration enables high accuracy with minimal calibration drift~\cite{sugano2015self}, supporting commercial implementations in platforms such as Apple Vision Pro, HTC Vive Pro, and Meta Quest Pro.

\textit{Desktop-based systems} typically fix cameras on top of stationary monitors. In these contexts, the user's head moves freely (6-DoF) relative to the stationary sensor, while eyes contribute an additional 3-DoF of rotational movement. Researchers have addressed this variability through head pose estimation methods~\cite{zhang18revisiting, lepetit2009ep}, physical constraints (e.g., chin rests), dual-camera setups~\cite{cheng2023dvgaze}, and online head-pose clustering~\cite{sugano2015appearance, sugano2014learning}.

\textit{Handheld mobile devices} present the most challenging scenarios for eye tracking due to unconstrained and independent movement of both device and user. This highly dynamic head-eye-camera relationship comprises up to 15-DoF: 6-DoF from head (three translational, three rotational), 6-DoF from device during hand holding, and 3-DoF from eyes. Despite extensive research efforts, including large-scale datasets such as GazeCapture~\cite{krafka2016eye}, TabletGaze~\cite{huang2017tabletgaze}, and RGBDGaze~\cite{arakawa2022rgbdgaze}, mobile systems remain particularly vulnerable to domain shifts induced by postural changes.

\subsection{Gaze Estimation Techniques}
Gaze estimation methodology broadly divides into model-based and appearance-based approaches~\cite{duchowski2017eye, hansen2009eye, cheng2024benchmark, lei2023end}. \textit{Model-based methods} rely on geometric eye models and optical principles, utilizing specialized hardware such as infrared illuminators to capture corneal reflections and pupil centers~\cite{duchowski2017eye}. While achieving high accuracy under controlled conditions~\cite{tonsen2017invisibleeye}, these methods are limited by their dependence on structured lighting and calibrated hardware, restricting deployment across commodity devices.

\textit{Appearance-based methods} employ data-driven techniques to learn the mapping from input images to gaze direction or points without explicit geometric modelling. These approaches leverage deep neural networks, such as CNNs~\cite{krafka2016eye, bao2021adaptive, valliappan2020accelerating, Huynh2022imon} and Vision Transformers (ViT)~\cite{cheng2022gaze}, demonstrating promising performance across diverse users and real-world environments~\cite{zhang2015appearance, cheng2022puregaze}. Appearance-based techniques can be further grouped into \textit{2D} and \textit{3D} gaze estimation methods.

\textit{2D gaze estimation} directly maps input images to screen coordinates~\cite{krafka2016eye, Huynh2022imon, valliappan2020accelerating}, offering computational efficiency for mobile platforms. However, these approaches often entangle head pose with eye movements, limiting generalizability across varying viewing conditions, particularly when motion introduces additional variability. \textit{3D gaze estimation} decouples head pose from eye rotation by predicting 3D gaze vectors in camera coordinate systems~\cite{zhang18revisiting}. While providing greater flexibility across hardware setups, these methods require precise camera-to-screen calibration and geometric modelling, presenting additional computational challenges.

In this work, we focus on 2D gaze estimation due to the prevalence of 2D mobile gaze estimation datasets such as GazeCapture~\cite{krafka2016eye} and RGBDGaze~\cite{arakawa2022rgbdgaze}. Several architectures have advanced this field in recent years. iTracker~\cite{krafka2016eye} serves as a foundational model for mobile environments, employing a multi-stream CNN architecture that processes face, eye regions, and face position grid to directly map visual features to screen coordinates. Valliappan et al.~\cite{valliappan2020accelerating} have proposed a computationally efficient approach using only eye patches as input, coupled with a lightweight SVR calibration model to enable on-device inference with reduced latency. AFF-Net~\cite{bao2021adaptive} is an adaptive feature fusion technique within a multi-stream CNN framework to leverage structural similarities between eyes and their relationship to facial landmarks, thereby enhancing prediction accuracy. 

More recent advancements include iMon~\cite{Huynh2022imon}, which incorporates an image enhancement module to refine visual details prior to gaze estimation and implements a specialized calibration scheme to address kappa angle errors. GazeTR~\cite{cheng2022gaze} applies vision transformer (ViT) as backbone into the gaze estimation domain and achieves better performance compared to CNN backbones. Driven by the superior performance of vision transformer~\cite{cheng2024benchmark}, we adapt the efficient-oriented MobileViT-v2~\cite{mehta2022separable} architecture to maintain transformer advantages while optimizing for mobile deployment.

\subsection{Calibration in Gaze Tracking}
Calibration is fundamental to gaze estimation accuracy across methodological approaches. Traditional eye trackers typically employ one-time static calibration procedures before system operation~\cite{duchowski2017eye, sugano2015self}, requiring users to fixate on predefined targets often arranged in a grid pattern. This establishes personalized mappings between unique eye characteristics and specific screen coordinates or gaze directions. However, such calibration inherently assumes relatively fixed head-eye-camera geometry, an assumption frequently violated in mobile scenarios~\cite{lei2025quantifying}.

To mitigate reliance on rigid calibration, researchers have explored adaptive~\cite{huang2016building, liu2024calibread, cheng2023dvgaze}, online~\cite{sugano2008incremental}, and calibration-free approaches. Adaptive techniques continuously update gaze models during use, leveraging implicit signals from user behaviours or contextual cues. For instance, Sugano et al.~\cite{sugano2015appearance} have employed mouse-click events as weak supervision for dynamic model updates, while Huang et al.~\cite{huang2016building} build personalized gaze models from interaction histories. Other approaches have utilized implicit saliency information~\cite{yang2021vgaze}, reading behaviours~\cite{liu2024calibread}, and touch interactions~\cite{cai2025gazeswipe} for calibration, though the latter cannot address scenarios without touch events.

In addition, research has explored low-complexity or few-shot calibration techniques~\cite{park2019few}. Chen and Shi~\cite{chen2020offset, chen2022towards} decompose gaze prediction into subject-independent components inferred from images and subject-dependent bias terms, enabling efficient calibration with minimal targets. Meta-learning approaches like FAZE~\cite{park2019few} train networks for rapid adaptation with limited samples, while generative methods using GANs~\cite{aranjuelo2024learning} augment sparse calibration datasets to improve robustness across diverse conditions.

However, few projects have looked into triggering calibration via IMU sensors nor applying continual learning to incrementally learn and maintain the knowledge on a diverse range of motion conditions over time to reduce the need to calibrate whenever there is a change. 

\subsection{Human Activity Recognition}
IMU sensors have become fundamental within mobile and wearable computing, enabling human activity recognition (HAR), context inference, and health monitoring~\cite{banjarey2021survey, gu2021survey}. IMU-based HAR can detect common activities; e.g., sitting, walking, lying down, using acceleration and angular velocity signals. HAR has evolved from traditional feature engineering with classical machine learning algorithms~\cite{bulling2014tutorial} to deep learning architectures, including CNNs~\cite{yang2015deep}, RNNs, hybrid CNN-LSTM models~\cite{zeng2014convolutional}, attention mechanisms~\cite{murahari2018attention, ek2022lightweight}, and transformers~\cite{luo2025bi}. These advancements have significantly improved performance across benchmark datasets such as WISDM~\cite{kwapisz2011activity, weiss2019wisdm} and HHAR~\cite{stisen2015smart}.

Within eye tracking, IMUs are underutilized. Some studies have employed IMUs for coarse-grained pose estimation~\cite{liu20243d, de2024combining} or eye movement prediction~\cite{satriawan2023predicting}, but few have integrated IMU-based HAR directly into gaze-tracking pipelines to enhance robustness under diverse motion conditions. This represents a critical research gap for mobile eye tracking in terms of how they challenge the precision of gaze tracking over time. 

Recent advancements in HAR have leveraged sophisticated deep learning architectures to improve recognition accuracy~\cite{gu2021survey, lara2012survey}. Singh et al.~\cite{singh2020deep} have incorporated self-attention mechanisms into sensor stream analysis, outperforming conventional approaches with 90.41\% accuracy on the WISDM dataset. This attention-based approach enables the model to focus on temporally significant segments within sensor data streams, enhancing feature discrimination.

Autoencoder-based architectures have emerged as effective solutions for handling noisy sensor streams in activity recognition tasks. Thakur et al.~\cite{thakur2022convae} combine convolutional autoencoders with LSTM networks for smartphone-based activity recognition, achieving 98.14\% accuracy on the UCI-HAR dataset with a 400-length sliding window. Similarly, An et al.~\cite{an2023supervised} develop an autoencoder with self-attention (SAE-SA) that reached 95.76\% accuracy on WISDM and 97.60\% on PAMAP2 datasets. These approaches demonstrate the robustness of autoencoder architectures in extracting effective representations from noisy IMU signals. 

Beyond supervised methods, unsupervised clustering approaches address challenges associated with manual sensor data annotation. Ma et al.~\cite{ma2021unsupervised} proposed a framework utilizing autoencoders for latent feature extraction followed by k-means clustering to generate pseudo-labels, achieving 55.0 Normalized Mutual Information (NMI) and 0.65 F1 score on HHAR dataset~\cite{stisen2015smart}. Extending this approach, Mahon et al.~\cite{mahon2023efficient} combine encoder-generated latent features with uniform manifold approximation for dimensionality reduction and Hidden Markov Models (HMM) for pseudo-label generation, improving performance to 67.9 NMI and 0.59 F1 score on HHAR. These unsupervised methods offer promising avenues for automatic motion pattern discovery without extensive manual annotation requirements.

Despite the significant potential of IMU sensors to complement visual signals in mobile gaze tracking, particularly for detecting calibration distortion induced by motion, existing adaptive and online methods~\cite{huang2016building, sugano2015appearance, zhong2024uncertainty} have not been extensively optimized for mobile settings.

\subsection{Research Gaps}
Recent work shows progress in gaze estimation, calibration, and human activity recognition. However, robust mobile gaze tracking in real-world interactions remains challenging. User motion and device handling create dynamic physical motion conditions that degrade mobile gaze tracking performance. This degradation, for example, impacts how users interact with on-screen content in daily applications, such as gaze-assisted reading~\cite{lei2023DynamicRead}, target selection~\cite{namnakani2023comparing, namnakani2025stretch, cai2025gazeswipe}, and gaze-hand multimodal inputs~\cite{kong2021eyemu}. We identify several research gaps related to these calibration issues:

First, how on-device motion sensors (IMUs) can properly trigger recalibration is underexplored. Current adaptive methods often fail to pinpoint the optimal recalibration moment using dynamic motion data. Second, systems must adapt to varied user motion. Traditional calibration often fails to learn new conditions without catastrophic forgetting. Robust continual learning strategies are thus needed to learn from diverse motion contexts without losing prior knowledge. Finally, few holistic frameworks integrate motion sensing with continual learning for mobile gaze tracking. Such an integrated approach offers the potential to maintain accuracy in real-world mobile use. Addressing these gaps is essential for practical mobile gaze tracking. Our work proposes a motion-aware continual calibration approach to tackle these challenges to enable gaze tracking robustness across realistic mobile environments.

\section{Proposed Approach}\label{sec:approach}
We introduce \textit{MAC-Gaze}, a continual calibration technique for mobile gaze tracking that leverages both visual and motion information to adapt to dynamic user conditions. Our approach integrates a backbone gaze estimation model with an IMU feature extractor for HAR and employs continual learning to update the calibrator as users transition between different activities. Our proposed system consists of the following components:
\begin{itemize}
    \item \textit{Backbone gaze estimation model} $M_V$: an appearance-based gaze estimation model that processes image input for point of gaze prediction and extracts latent visual features for the continual calibration module;
    \item \textit{Motion-aware Detector} $M_I$: a human activity recognition model that processes IMU sensor data to extract motion features representing different motion conditions and classifies broader user activities;
    \item \textit{Calibration Trigger}: A hybrid decision-making mechanism that determines precisely when to initiate recalibration of $M_C$. It utilizes supervised activity classifications from the \textit{Motion-aware Detector} ($M_I$) to identify significant changes in user state, complemented by unsupervised clustering techniques applied to raw IMU data to detect novel, fine-grained motion patterns not previously encountered;
    \item \textit{Continual Calibration Module} $M_C$: a calibrator that takes visual features from $M_V$ to produce personalised gaze predictions. $M_C$ is continually updated, when prompted by the \textit{Calibration Trigger}, to account for dynamic and diverse motion conditions. Compared to $M_V$, $M_C$ is typically a smaller machine learning model, trained on a much smaller, user- and device-specific gaze dataset.
\end{itemize}

\begin{figure}[!htbp]
    \centering
    \includegraphics[width=0.95\linewidth]{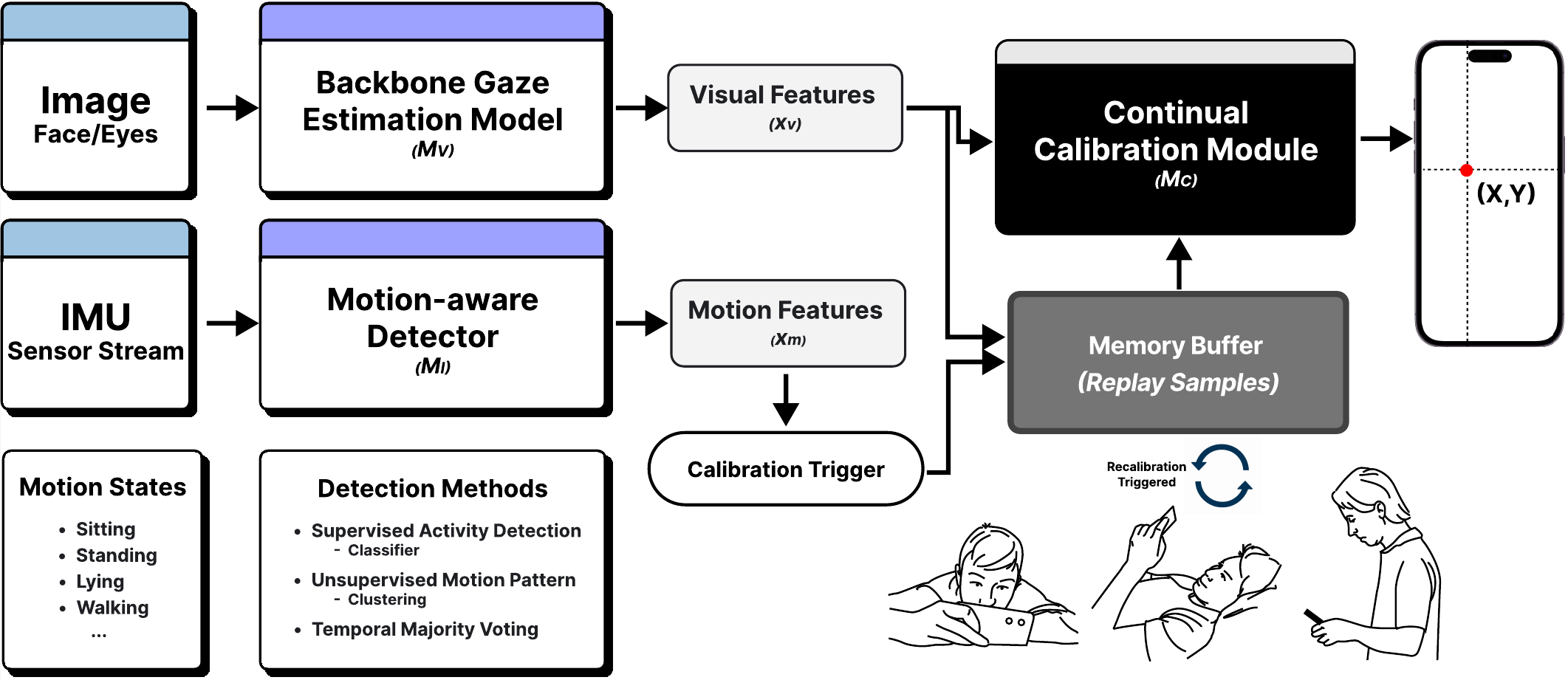}
    \caption{Workflow of Motion-Aware Continual Learning Calibration for Mobile Gaze Tracking}
    \label{fig:workflow}
\end{figure}

In general, $M_V$ can be trained on independent, large gaze datasets that consist of a large number of facial images from many participants and gaze points as labels. $M_V$ forms the backbone model and it often leads to low accuracy of gaze prediction when directly applied to a target user. To tackle the problem, a gaze-tracking system launches the following \textit{calibration} process. It collects a small set of gaze data on the target user, including facial images and gaze ground truth. The facial images will be passed through $M_V$ to generate visual features, which along with gaze ground truth will be used to train a machine learning model $M_C$ to better adjust gaze prediction. In our system, the key novelty is to use IMU data in a mobile device, processed by $M_I$ and the \textit{Calibration Trigger}, to detect changes in motion states, launch the calibration when necessary, and continuously update the calibrator $M_C$ to maintain accuracy across various motion states. The overview of the system workflow is presented in Figure~\ref{fig:workflow} and we will illustrate each component in more details.

\subsection{Backbone gaze estimation model}
The backbone gaze estimation model ($M_V$) serves as the foundation for our motion-aware continual calibration framework, extracting discriminative features from visual inputs. Following state-of-the-art methodologies, we first detect and crop the face and eye regions via Google's ML kit, resizing them to $224\times 224$ and $112\times 112$, respectively. We also generate a $25 \times 25$ binary mask via Dlib~\cite{king2009dlib} as a face grid, which indicates the head position in the original image and provides information about the distance between the face and camera. These facial and eye patches, along with the face grid, are subsequently processed through a deep neural network for gaze estimation.

We adopt the MobileViT-v2 architecture~\cite{mehta2022separable} to extract 256-dimensional visual features independently from three input images: the full face and two eye patches. MobileViT-v2 is chosen for its efficient architecture, which combines the lightweight characteristics of MobileNet with the global context modelling capabilities of Vision Transformers. This balance between computational efficiency and representational power makes it possible to deploy our mobile gaze tracking system to mobile devices in the future. 

The face grid is processed through two fully connected layers with 256 and 128 neurons, respectively. The resulting feature vectors, consisting of three from visual inputs (256*3) and one from face grid (128), are concatenated into a single vector of 896 dimensions. This combined vector will be passed through three fully connected layers (896 $\rightarrow$ 256 $\rightarrow$ 128 $\rightarrow$ 2) to regress the 2D gaze point. 

We use the MobileViT-v2 model pre-trained on ImageNet from the timm library~\cite{rw2019timm} and finetune it on the GazeCapture dataset~\cite{krafka2016eye}, containing 1.4M gaze-annotated images collected on mobile devices. After training, we freeze the model weights for directly inference and use the output from the third-to-last layer as visual features for the calibrator.

\subsection{Motion-aware Detector}
In mobile gaze estimation, accuracy is significantly impacted by user movements and varying phone holding postures~\cite{lei2023DynamicRead}, leading to performance degradation. To address this challenge, our system incorporates a dedicated \textit{Motion-aware Detector ($M_I$)}. This component processes data from the device's IMU sensors, such as the accelerometer and gyroscope, capture the phone's dynamic movement and rotation. The core function of $M_I$ is to learn robust motion features from these IMU signals and to classify the user's broader activity or current motion state. The insights derived from $M_I$, specifically the extracted features and activity classifications, are subsequently utilized by the \textit{Calibration Trigger} (detailed in Section~\ref{subsec:calibration_trigger}) to inform decisions about when automatic recalibration of the gaze model is necessary.

Building upon the recent literature in HAR in Section~\ref{sec:related}, we employ an encoder-decoder architecture with an additional classification head. Specifically, we adopt Supervised Autoencoder with Self-Attention (SAE-SA)~\cite{an2023supervised} and enhance it with Squeeze-and-Excitation blocks~\cite{hu2018squeeze} for more effective IMU feature extraction in short sliding window and fewer sensors. The encoder begins with a 1D convolutional layer (kernel size of 7) that extracts initial temporal features from the accelerometer signals. This is followed by three residual blocks with increasing dilation rates (1 $\rightarrow$ 2 $\rightarrow$ 4) and channel dimensions (64 $\rightarrow$ 128 $\rightarrow$ 256) that capture both local and increasingly global motion patterns. Each residual block incorporates SE attention that adaptively recalibrates channel-wise feature responses, emphasizing relevant motion components. A self-attention module is then applied to model temporal dependencies across the entire sequence, allowing the network to correlate motion patterns regardless of their temporal distance.

The encoded latent representation $\mathbf{z}_m$ is processed through two parallel pathways: (1) a classification branch with fully connected layers (256 $\rightarrow$ 512) that predicts the activity class, and (2) a reconstruction branch that reconstructs the input signal through transposed convolutions. This dual-task learning approach forces the model to learn meaningful representations that preserve both discriminative features for classification and structural information for reconstruction. During training, we employ a weighted loss function that balances reconstruction error ($r=0.3$) and classification accuracy ($1- r=0.7$).

\begin{align*}
\mathbf{z}_m &= \text{Encode}(\mathbf{x}_m), \quad \hat{\mathbf{x}}_m = \text{Decode}(\mathbf{z}_m), \quad \hat{\mathbf{y}}_{\text{activity}} = \text{Classify}(\mathbf{z}_m), \\
L &= \frac{1}{|D|} \sum_{(\mathbf{x}_m, \mathbf{y}_{\text{activity}}) \in D} \left[ r \cdot \|\mathbf{x}_m - \hat{\mathbf{x}}_m\|_2^2 + (1-r) \cdot \text{CE}(\mathbf{y}_{\text{activity}}, \hat{\mathbf{y}}_{\text{activity}}) \right],
\end{align*}
where $\mathbf{x}_m \in \mathbf{R}^{T\times M}$ represents the IMU input,  where $T$ is the temporal window length (e.g., 200 frames) and 
$M$ is the number of IMU measurement dimensions (e.g., 3-axes accelerometer readings). $D$ represents the training dataset, $\text{CE}$ denotes cross-entropy loss, and $r$ balances reconstruction and classification objectives. 

\subsection{Continual Calibration}
Adapted from a class-incremental learning problem~\cite{boschini2022class}, we first formulate continual calibration as follows. The model $M_C$ undergoes training with a sequence of tasks $T=[t_1, t_2, ..., t_N]$, where each task $t_i (i \in [1,N])$ corresponds to a continuous time-series gaze dataset $D_i = \{(\mathbf{x}^{(i,j)}_v, \mathbf{x}^{(i,j)}_m, \mathbf{x}^{(i,j)})\}_{j=1}^{J}$. Here, $\mathbf{x}_v^{(i,j)}$ represents facial images of a certain resolution, while $\mathbf{x}_m^{(i,j)} \in \mathbf{R}^{T\times M}$ represents IMU sensor data in dimension of $M$ collected over a window length of $T$. $y^{(i,j)}$ denotes the 2D gaze point coordinates as ground truth.  Unlike traditional tasks having different sets of classes, each $D_i$ is characterized by a unique distribution of motion features representing a different activity or device holding posture. The objective is to accurately predict gaze for all data whose motion characteristics align with the data up to task $i$.

A critical challenge in this process is to prevent catastrophic forgetting (CF); that is, the tendency of neural networks to lose performance on previously learned tasks when fine-tuned on new data. In our context, naively fine-tuning $M_C$ on new calibration data would optimize it for the current motion state but potentially degrade performance when the user returns to previously encountered postures. To address CF, three types of continual learning techniques have been identified~\cite{wang2024cl, de2021continual}. \textit{Regularization}-based techniques incorporate regularisation terms to penalise large updates to model weights. \textit{Replay}-based techniques select a small subset of informative samples from previous tasks and leverage them to update the model with new task data. This helps reinforce the learned knowledge and prevent the degradation of performance on earlier tasks. \textit{Dynamic architecture}-based techniques extend the model's architecture to increase its capacity of learning more knowledge. A recent survey~\cite{jha2021cl} shows that replay-based techniques can work well for a relative small number of tasks while maintaining low computational cost.  

In our setting, we adopt a replay approach, storing a representative subset of examples from previously encountered motion states in a memory buffer; that is, $\mathcal{B} = \{(x^{(i)}_v, x^{(i)}_m, y^{(i)})\}_{i=1}^{i=N_B}$, where $N_B$ is the size of the memory buffer and each entry $(x^{(i)}_v, x^{(i)}_m, y^{(i))})$ is downsampled from the observed tasks' datasets. When updating $M_C$, we use the buffer data $\{(x^{(i)}_v, y^{(i)}\}_{i=1}^{i=N_B}$ and the current calibration data $\{(x^{(i)}_v, y^{(i)})\}_{i=1}^{i=J}$, where facial images $x^{(i)}_v$ will be fed to the backbone gaze estimation model $M_V$ to extract visual features, as input to the calibrator $M_C$.  $M_C$ optimizes a combined loss function:

\begin{equation}
    L_C = \sum_{j=1}^{J} L_{\text{calibration}}(y^{(j)}, \hat{y}^{(j)}) + \alpha \sum_{k=1}^{N_B} L_{\text{replay}}(y_{\text{buffer}}^{(k)}, \hat{y}_{\text{buffer}}^{(k)}),
\label{eq:continual_loss}
\end{equation}
where $L_{\text{calibration}}$ represents the error on new calibration data, $L_{\text{replay}}$ maintains performance on replay samples from previous tasks, and $\alpha$ balances adaptation versus retention. This approach enables $M_C$ to progressively adapt to diverse motion contexts while maintaining accuracy across the full range of user activities.

\subsection{Calibration Trigger}\label{subsec:calibration_trigger}
A key challenge in continual calibration is determining \textit{when} recalibration is needed. Intuitively, we can start with a supervised approach where our HAR model $M_I$ is trained in a supervised manner on an independent HAR dataset covering a wide range of activity classes. If $M_I$ detects a new activity that has not been observed from previous tasks, then we trigger the calibration process. However, this approach might not be able to capture fine-grained device-holding postures when the user stays with the same activity. On the other hand, we can adopt an unsupervised approach which characterises motion features and assesses whether the current motion features significantly deviate from the observed features. This approach can potentially capture finer-grained posture variance, but its performance is generally lower than the supervised techniques.

To leverage the advantages of both supervised and unsupervised approach, 
we design a hybrid motion-aware calibration strategy that combines higher accuracy of supervised classification with the adaptability of unsupervised clustering. This approach 
where we first use the supervised classifier to detect significant motion transitions, and then apply unsupervised clustering to determine whether the new motion pattern warrants calibration.

\noindent\textit{Supervised activity recognition}: We use the classification head of SEAR-Net to identify transitions between predefined activities (e.g., sitting to standing to sitting). To prevent spurious triggering from momentary misclassification, we implement a temporal majority voting algorithm where to assess the consistency score~\cite{xia2024self} in a temporal window.

Let a window of raw IMU data accumulated over a consensus window size of $c$ up to a timestamp $t$ be $\mathbf{X}_t = [\mathbf{x}_r^{t-c+1}, \ldots, \mathbf{x}_r^t]$, where $\mathbf{x}_r^{j} \in \mathbf{R}^{1\times M}$ ($j \in [t-c+1, t]$). The above supervised classifier will perform activity recognition over each $T$-length IMU input $\mathbf{x}_m^j = [\mathbf{x}_r^{j},\mathbf{x}_r^{j+1}, ..., \mathbf{x}_r^{j+T-1}]$ and derive activity predictions $\mathbf{Y}_t = [\hat{y}^{t-c+1}, ..., \hat{y}^t]$, where $\hat{y}_j = Classify(Encoder(\mathbf{x}_m^j))$. Then a majority voting is conducted on $\mathbf{Y}_t$ to find the most frequently inferred activity labels and see if the frequency is above a certain threshold so as to be considered as a stable classification:
\begin{align*}
 \frac{|\{\forall y \in \mathbf{Y}_t : y = \text{mode}(\mathbf{Y}_t)\}|}{|\mathbf{Y}_t|} \geq \tau,
\end{align*}
where mode$(\mathbf{Y}_t)$ is the function to find the most frequent element in an array and $\tau$ is the consensus ratio threshold. If the inferred activity is stable and different from the current activity, then we move to the next step for fine-grained motion pattern recognition.

\noindent\textit{Fine-grained Motion Pattern Discovery via Clustering}
Our next objective is to determine whether the incoming sequence of IMU readings constitutes a novel, fine-grained motion pattern not adequately captured by the broader activity class. To this end, we maintain a compact memory buffer containing representative raw IMU samples from previously encountered tasks. These stored readings are then clustered using a Gaussian Mixture Model (GMM) to characterise the distribution of known fine-grained motion patterns.

\begin{figure}[!htbp]
    \centering
    \begin{subfigure}[b]{0.45\linewidth}
        \includegraphics[width=\linewidth]{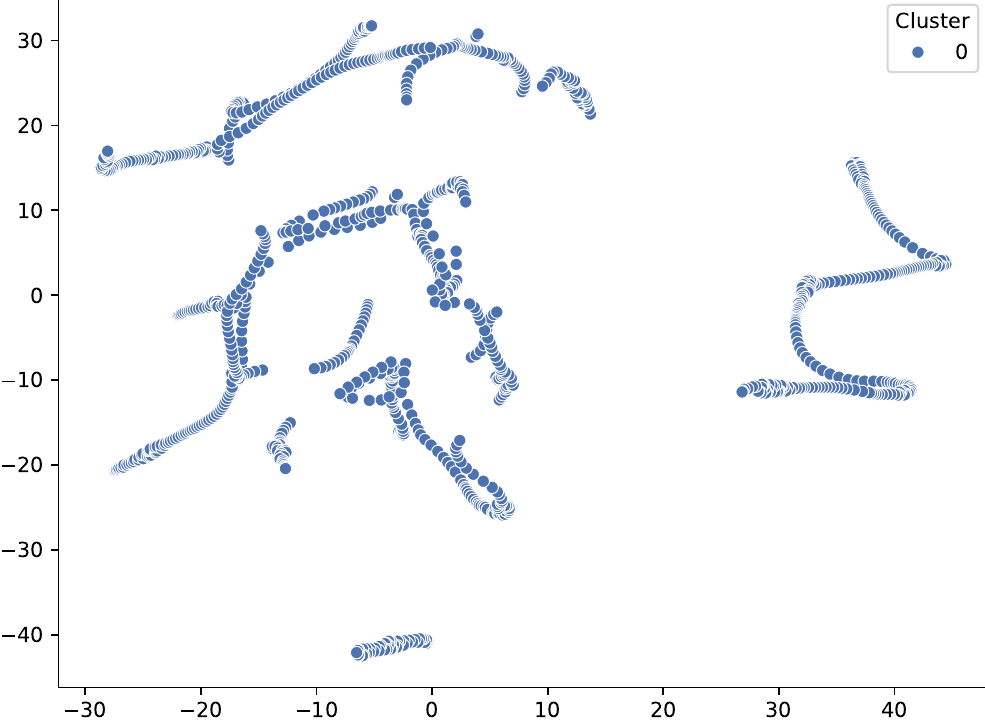}
        \caption{GMM clustering on latent features from SEAR-Net}
        \label{fig:subfig_latent_feature}
    \end{subfigure}
    \hfill
    \begin{subfigure}[b]{0.45\linewidth}
        \includegraphics[width=\linewidth]{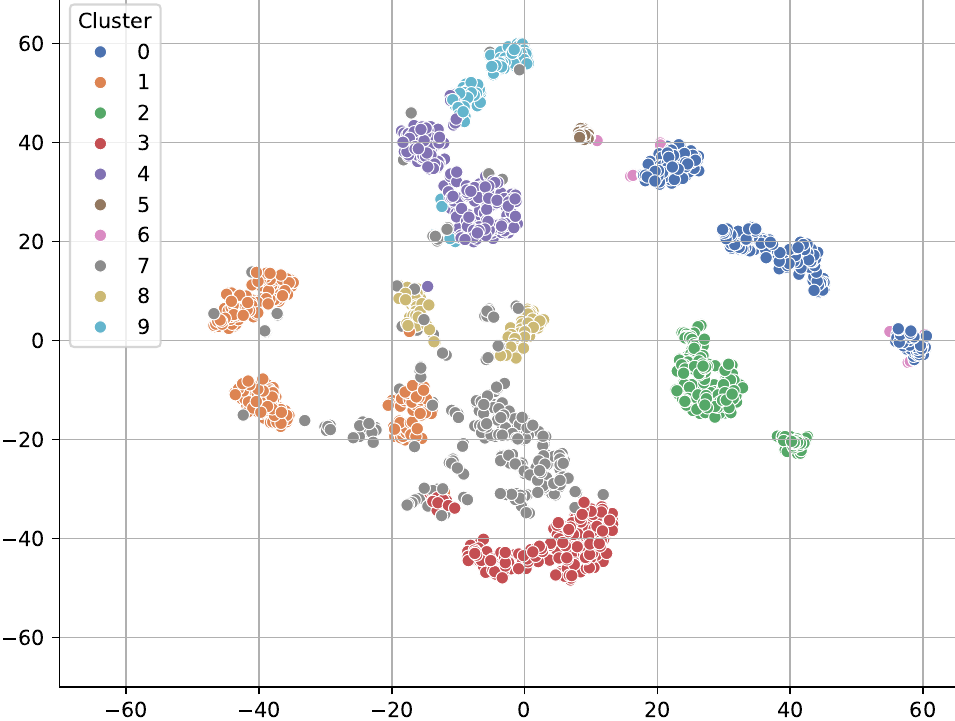}
        \caption{GMM clustering on raw IMU sensor readings}
        \label{fig:subfig_raw_imu}
    \end{subfigure}
    \caption{Comparison of feature representations for GMM clustering of fine-grained motion patterns, visualized via t-SNE using a series of calibration sets from Participant 11. (a) t-SNE visualization of latent features extracted by SEAR-Net, showing poor cluster separation with only one identified cluster. (b) t-SNE visualization of raw IMU sensor readings, revealing distinct and well-separated clusters.}
    \label{fig:clustering_comparison}
\end{figure}

Initially, we considered using the latent features ($\mathbf{z}_m$) extracted by our SEAR-Net IMU encoder for clustering fine-grained motion patterns. The SEAR-Net encoder is optimized to produce representations that are discriminative for broad activity classes (e.g., sitting, walking) over 200-sample windows. However, this optimization process, which includes convolutional layers and self-attention mechanisms designed to capture class-relevant temporal dependencies, inherently leads to a degree of feature smoothing. As visualized using t-SNE, Figure~\ref{fig:subfig_latent_feature}, these window-level latent features formed a dense, largely overlapping cluster, from which GMM struggled to discern more than one or two indistinct motion states within a broader activity class. This suggests that the subtle, rapid postural shifts and micro-movements critical for triggering fine-grained recalibration may be averaged out or abstracted away in these higher-level latent representations.

To capture these finer variations, we opted to perform GMM clustering directly on the individual raw IMU measurement vectors ($\mathbf{x}_r \in \mathbb{R}^{1\times M}$ from the incoming sequence). As shown in Figure~\ref{fig:subfig_raw_imu}, this approach yielded a clearer separation into multiple distinct clusters (e.g., 10 clusters identified). While raw IMU data is inherently noisier and more susceptible to spurious changes, potentially leading to over-triggering, we mitigate this through the temporal majority voting mechanism for broad activity changes and careful threshold selection for outlier detection in the GMM likelihood stage. The direct use of raw IMU data for fine-grained pattern discovery, despite its noise sensitivity, was thus deemed more effective for detecting subtle deviations that might necessitate recalibration, which might otherwise be missed by clustering on the smoothed latent features designed for broader activity classification. Future work could explore hierarchical clustering approaches or modified encoder architectures that better preserve both coarse and fine-grained motion details.

The new motion pattern detection is to assess whether the majority of collected new data in $X_t$ conforms to the probability density estimated by the GMM learned on the buffer data $\mathcal{B}$. That is, $\forall x_r \in \mathbf{X}_t$, we evaluate its highest likelihood of belonging to any component of the GMM:
\begin{align*}
    p_{r} \;=\; \max_{k} \Bigl(\pi_{k}\,\mathcal N(x_{r}\mid\mu_{k},\Sigma_{k})\Bigr),
\end{align*}
where $\{\pi_{k},\mu_{k},\Sigma_{k}\}_{k=1}^{K}$ are the mixture weights, means, and covariances for each of the $K$ Gaussian components (indexed by $k$) learned from the stored data.

If the likelihood is less than a threshold $\tau_i$ (e.g., 0.95), then we consider $x_r$ as an outlier. We calculate the outlier ratio among $\mathbf{X}_t$ as 
\begin{align*}
    r_t = \frac{|\{i \mid p_r(i) < \tau_i\}|}{c}.
\end{align*}
If the outlier ratio exceeds a threshold (e.g., 0.75), we conclude that the majority of $X_t$ does not conform to any previously observed motion pattern. Therefore, $\mathbf{X}_t$ is likely to correspond to a new, fine-grained motion pattern and trigger recalibration. Then we will downsample $\mathbf{X}_t$ along with their corresponding visual and ground truth data and merge with the buffer $\mathcal{B}$, upon which we will run GMM on the IMU input for future new motion pattern discovery. The specific values for these various thresholds and parameters within the hybrid triggering strategy (e.g., temporal majority voting window size $c$, GMM component selection, likelihood threshold $\tau_i$, and outlier ratio $r_t$) were determined empirically based on performance on a held-out validation subset of our training data, aiming to balance trigger sensitivity with the prevention of excessive recalibrations.

\section{Experiment Setup}
The main objective of our work is to present a continual calibration framework that can automatically trigger recalibration and allows incremental calibration to adapt to a wide range and speed move of head-eye-camera variations. Towards this objective, we design the following experiment and evaluation methodology.

\subsection{Datasets}
In our system, the backbone gaze estimation model $M_V$ and activity recognition model $M_I$ are pre-trained on the publicly available datasets.

\subsubsection{HAR Datasets for Training $M_I$}
There are various HAR datasets available and our selection criteria is that (1) the dataset is collected on smart phones or watches; (2) it includes activities similar to our motion conditions; and (3) it contains a large set of samples to learn effective motion features. Based on these criteria, we have selected Heterogeneity Human Activity Recognition (HHAR) dataset~\cite{hhar2015data}, which contains accelerometer and gyroscope data from smartphones and smartwatches across six activities: biking, sitting, standing, walking, stair up, and stair down. HHAR is collected from 9 participants and totally including 14M samples, and 11M from phone and 3M from watch. We only use the phone data to train our supervised IMU model $M_I$; that is, 1.8M in biking, 2M in sitting, 1.6M in stair down, 1.78M in stair up, 1.8M in standing and 2.2M in walking.

\subsubsection{Motion-aware Gaze Datasets for Calibration (MotionGaze)}
Our continual calibration technique requires the mobile gaze datasets that have collected continuous IMU sensor readings along with facial images and gaze ground truth, ideally across multiple motion conditions. As the use of IMU in mobile gaze estimation is still a new topic, there only exist a few datasets; e.g., \textit{RGBDGaze}~\cite{arakawa2022rgbdgaze}. This dataset contains synchronized RGB images, depth maps, and IMU sensor data from 45 participants performing gaze estimation tasks in four distinct postures: sitting, standing, lying down, and walking. After our initial investigation, the RGBDGaze dataset exhibits the following limitations: (1) no natural transitions between postures, (2) restricted device handling, and (3) a much narrow range of dynamic movements. To tackle these limitations, we collect our own dataset to better reflect more natural, varied motion conditions.

Our dataset, called \textit{MotionGaze}~\cite{lei2025quantifying}, was collected from 10 participants (4 females, 6 males, $M=26.4$ years, $SD = 3.14$) using our prototype mobile gaze tracking system~\cite{lei2025quantifying} within a large laboratory room (approximately 8m $\times$ 20m) configured for various activities. Participants performed tasks under five distinct motion conditions designed to simulate common daily scenarios: lying down, sitting, standing, walking in an open space, and walking while navigating a maze. Our data collection protocol aimed to capture not only the obvious transitions between these postures but also the natural micro-motions, subtle postural adjustments, and variations in device handling (e.g., subtle device movements occurring when shifting tasks, such as from a video call to news Browse, or minor adjustments in device angle) that users exhibit even when predominantly in one of these states; these subtle dynamics are crucial as they contribute to the overall 'motion' profile from a sensor perspective. For the "walking in an open space" condition, participants walked at their natural, self-selected pace over varying distances within a designated clear area of the room (approximately 8m $\times$ 10m), allowing for natural gait patterns and device sway. To elicit more complex and varied ambulatory movements, the "walking while navigating a maze" condition required participants to traverse a 5m $\times$ 10m concave-shaped maze, whose walking path was enclosed by soft cushions. In each condition, they completed two gaze tasks: (1) a pursuit task, in which they followed a dot moving along the screen’s boundary for 5 seconds per session, resulting in approximately 125–150 ground truth gaze points; and (2) a fixation task, where participants fixated on a sequence of 9 dots appearing for 3 seconds each. The dots were uniformly distributed across the screen, and their presentation order was randomized. Each fixation session lasted 27 seconds, producing approximately 675–810 labelled gaze points. The participants were asked to perform these tasks as naturally as possible. This dataset, therefore, captures a wide range of real-world usage patterns, including natural variations in posture, diverse continuous and transitional movements, and varied device handling during gaze-based interaction. Consequently, it is structured to support detailed analyses of how gaze behavior is affected by, and can be adapted to, both continuous dynamic activities (like walking) and periods within, or transitions across, different, more stable postures. Each participant’s experiment took approximately 1.5 hours to complete, totally collecting 481K images with 803K IMU readings. Given the time and effort involved in conducting the study, we limited the sample to 10 participants. We use the data of the task (1) for training and calibration and the data of the task (2) for testing.

\subsubsection{Dataset Comparison} 
\begin{figure}[!htbp]
    \centering
    \includegraphics[width=0.95\linewidth]{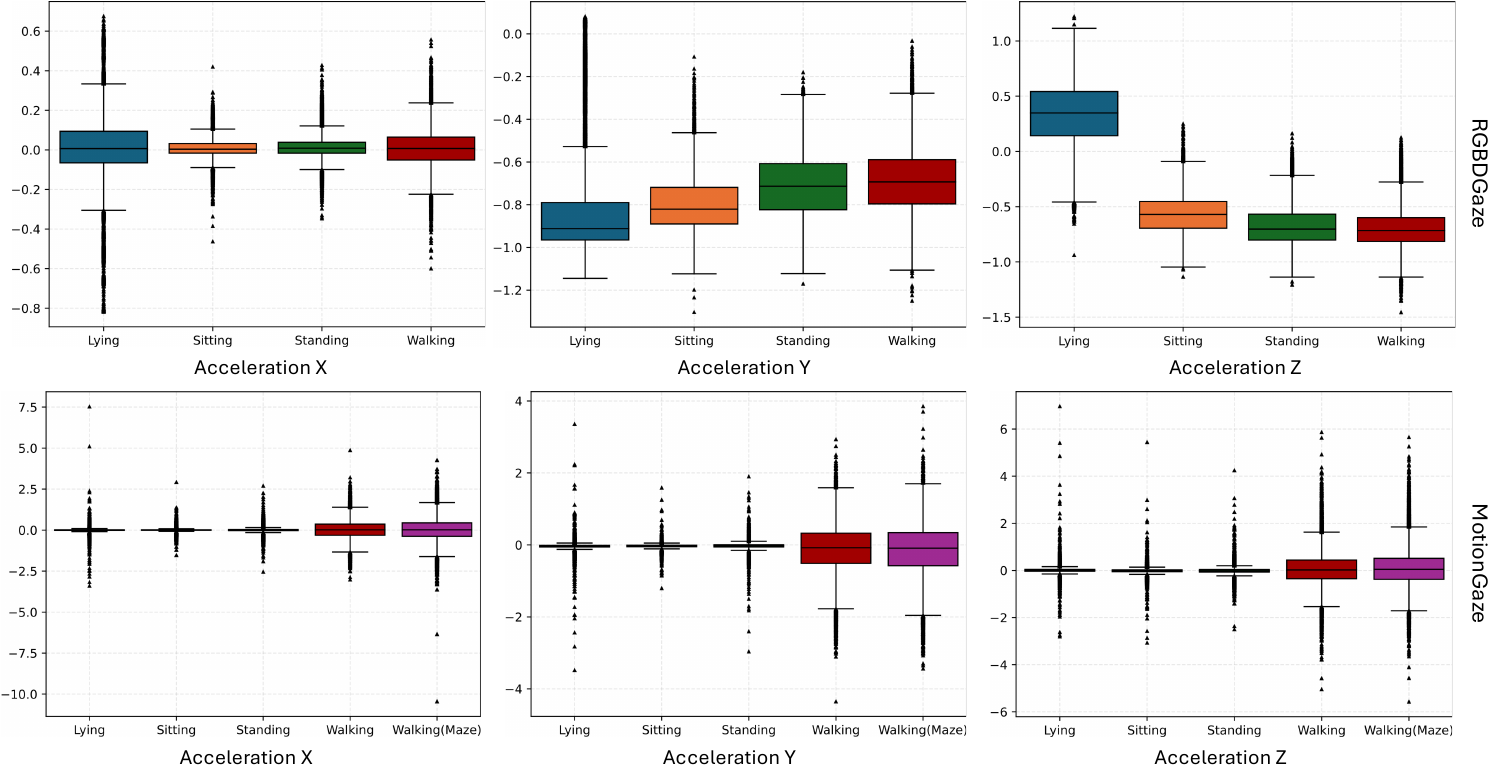}
    \caption{Comparison of acceleration ranges between RGBDGaze and MotionGaze in box plots}
    \label{fig:datasets_acceleration_boxplots}
\end{figure}

To better understand motion patterns in both datasets, we examined the distribution of tri-axial accelerometer readings (X, Y, Z) across postural conditions. As shown in Figure~\ref{fig:datasets_acceleration_boxplots}, the RGBDGaze dataset exhibits relatively constrained acceleration ranges across all axes, with walking and standing showing modest variability and lying showing tight distributions particularly in the Z-axis, where values remain near $0.5 \sim 1.0$ g. In contrast, MotionGaze demonstrates significantly broader acceleration ranges, especially under dynamic conditions. More specifically, walking and walking in maze produce larger variances and outliers, with Z-axis values spanning approximately from $-2$ g to over $6$ g.

To quantitatively validate these observed differences, we conducted statistical analyses comparing the accelerometer data between the two datasets. One-way ANOVA tests revealed highly significant differences across all three axes: X-axis ($F = 59.70$, $p < 0.0001$), Y-axis ($F = 300821.80$, $p < 0.0001$), and Z-axis ($F = 68288.30$, $p < 0.0001$). Subsequent t-tests confirmed these disparities with large effect sizes: X-axis ($t = -8.74$, $p < 0.0001$), Y-axis ($t = -604.64$, $p < 0.0001$), and Z-axis ($t = -262.59$, $p < 0.0001$). These statistical findings provide evidence that our MotionGaze dataset captures significantly more diverse motion conditions than RGBDGaze, particularly in the Y and Z dimensions where the differences are the largest. Such variability emphasizes the importance of motion-aware calibration techniques for robust mobile gaze tracking in unconstrained, real-world settings.

\subsection{Evaluation Metrics}

We evaluate the gaze estimation accuracy using the Euclidean distance between the predicted and true gaze points:
\begin{equation} 
\text{Error}_{\text{2D}} = \frac{1}{N} \sum_{i=1}^{N} || \mathbf{p}_i - \hat{\mathbf{p}}_i ||, 
\end{equation}
where $\mathbf{p}_i$ is the true gaze point and $\hat{\mathbf{p}}_i$ is the predicted gaze point.

\subsection{Evaluation Procedures}
To comprehensively evaluate the performance of our continual calibration approach, we designed systematic procedures for both the RGBDGaze dataset and our MotionGaze dataset. These procedures were designed to assess how well our system adapts to different posture transitions and evolves over time as it encounters new motion contexts.

For the backbone gaze estimation model $M_V$ and the HAR model $M_I$, we initially trained using publicly available datasets: GazeCapture for gaze estimation and HHAR for HAR. We then fine-tuned the models on smaller subsets of the RGBDGaze and MotionGaze datasets, respectively, to adapt it to the specific device and conditions of each target dataset prior to its use as a feature extractor for calibration. This fine-tuning step is necessary because the visual and motion data distributions in RGBDGaze and MotionGaze differ significantly from those in GazeCapture and HHAR, making adaptation essential for achieving reasonable performance.
To ensure subject independence and realistic evaluation, we split both RGBDGaze and MotionGaze datasets based on participants. Specifically, for RGBDGaze, we used data from 9 participants for training and 36 for testing. For MotionGaze, we used 3 participants for training and 7 for testing.

The calibration model's performance was evaluated using data from each participant in the test sets of both datasets. We defined each motion condition (e.g., lying, sitting, walking) as a separate task and generated multiple evaluation sequences by systematically varying the initial task, ensuring each condition served as a starting point (e.g., Lying $\rightarrow$ Sitting $\rightarrow$ Standing $\rightarrow$ Walking; Sitting $\rightarrow$ Standing $\rightarrow$ Walking $\rightarrow$ Lying; etc.). The underlying data for these evaluations was gathered with participants consistently instructed to maintain a natural vertical phone grip and screen visibility, which allowed this approach to assess the impact of different starting points and transition orders on calibration performance.

For each participant, we first initialized our continual calibrator using 10\% of the data (approximately 90-105 frames) from their initial posture. We then tested the calibrated model on the remaining 90\% of data from all the motions. When our motion detector identified a change in posture, we triggered recalibration using 10\% of the data from the newly detected motion. After each recalibration, we evaluated performance across all motion conditions to assess how the calibrator performs on all the observed conditions. This process continued until all conditions in the sequence had been processed. We repeated this procedure for all the test participants and averaged the results to ensure robust evaluation.

\subsection{Implementation Details}
We implemented our system using PyTorch 2.4.1, CUDA 12.4, AMD EPYC 7282, 256 GB RAM, and NVIDIA A100 GPU.

\subsubsection{Backbone Gaze Estimation Model}
The backbone gaze estimation model ($M_V$) was trained using hyperparameters including a batch size of 256, an initial $10^{-4}$ learning rate, L1 loss, 0.1 weight decay, and up to 120 epochs with early stopping. After its initial pre-training on the GazeCapture dataset, $M_V$ achieved a 1.57 cm mean error on the GazeCapture phone test set. The trained model then yielded a 2.45 cm mean error when directly evaluated on the RGBDGaze  dataset, as shown in Table~\ref{tab:domain_finetune} under "Direct". Further fine-tuning $M_V$ on a 9-participant subset of RGBDGaze (from its 45 total participants) reduced the mean error on the remaining RGBDGaze test participants to 1.76 cm overall, with all motion conditions achieving errors below 2 cm, under "Fine-tuned" for RGBDGaze in Table~\ref{tab:domain_finetune}. This adapted $M_V$ was thus considered suitable as a base model and feature extractor for the subsequent continual calibration evaluations.

\begin{table}[!htbp]
\centering
\resizebox{0.65\textwidth}{!}{
\begin{tabular}{cccc} 
\toprule
\multirow{2}{*}{\textbf{Motion}} &\textbf{GazeCapture (Phone)}  & \multicolumn{2}{c}{\textbf{RGBDGaze (Phone)}} \\ 
\cmidrule{3-4}
 &  & Direct & Fine-tuned  \\ 
\cmidrule{1-4}
Lying      & --   & 2.17 & 1.80 $\downarrow$  \\
Sitting   & --   & 2.36 & 1.76  $\downarrow$ \\
Standing   & --   & 2.41 & 1.70  $\downarrow$ \\
Walking    & --   & 2.42 & 1.77 $\downarrow$  \\
\cmidrule{1-4}
Overall    & 1.57 & 2.45 & 1.76  $\downarrow$ \\
\bottomrule
\end{tabular}}
\caption{Mean Euclidean errors (cm) of the $M_V$ gaze estimation model on GazeCapture (Phone) and RGBDGaze (Phone) datasets. 'Direct' denotes testing on the full RGBDGaze dataset; 'Fine-tune' denotes results after fine-tuning on a 9-participant subset and testing on the remaining 36.}
\label{tab:domain_finetune}
\end{table}

\subsubsection{HAR Model Implementation and Selection}
For SEAR-Net's implementation and optimization, we first preprocess the IMU sensor data by segmenting it into sliding windows of 200 timestamps with 75\% overlap and applying z-normalization to each sensor channel for standardization. Through extensive grid search for hyperparameter tuning, we determined the optimal architecture configuration: ReLU activation functions with dropout (rate=0.1) between convolutional layers to mitigate overfitting. The training process utilized the Adam optimizer with an initial learning rate of $1 \times 10^{-4}$ and a learning rate scheduler that reduced the rate by a factor of 0.1 when validation loss plateaued for 10 epochs. We use a batch size of 256, and 200 epochs. Additionally, we implemented an early stopping strategy (patience=10, min\_delta=0.001) to prevent overfitting, preserving the model weights that achieved optimal performance on the validation set.

To make sure the performance of our HAR models good enough for motion detection, we perform the following evaluation. We evaluate the activity recognition and clustering performance on both HHAR~\cite{stisen2015smart} and RGBDGaze. For the classification task, we measure the performance in F1-scores and compare with two commonly used baselines such as 1DCNN~\cite{zeng2014convolutional} and SAE-SA~\cite{an2023supervised}. We also compare with clustering techniques such as Deep Clustering~\cite{mahon2023efficient} and VaDE (variational deep embedding)~\cite{jiang2016variational} to see if they can reliably detect motion patterns. For the clustering task, we employ normalized mutual information (NMI) to measure the agreement between the true and predicted clusters with normalization to account for different sizes of clusters.

\begin{equation}
    \text{NMI}(Y, C) = \frac{2 \times I(Y; C)}{H(Y) + H(C)}
\end{equation}
where $I(Y; C)$ is the mutual information between true labels $Y$ and predicted clusters $C$, and $H(\cdot)$ is the entropy.

\begin{table}[!htbp]
\centering
\resizebox{0.85\textwidth}{!}{
\begin{tabular}{lccclcc} 
\toprule
\multirow{2}{*}{\textbf{Dataset}} & \multicolumn{3}{c}{\textbf{Supervised Classifier}} &  & \multicolumn{2}{c}{\textbf{Unsupervised Clustering}}  \\ 
\cmidrule{2-4}\cmidrule{6-7} 
& 1DCNN & SAE-SA & \textbf{SEAR-NET(Ours)}  &  & VaDE-HAR    & {DeepClustering} \\ 
\cmidrule{1-4}\cmidrule{6-7}
HHAR  & 0.73  & 0.89   & 0.93   &  & 0.44 (0.10) & 0.47 (0.10)  \\
RGBDGaze  & 0.64  & 0.77   & 0.80  &  & 0.23 (0.07) & 0.26 (0.08) \\
\bottomrule
\end{tabular}}
\caption{Motion detection performance (F1-scores, with NMI in brackets for clustering methods) on HHAR and RGBDGaze datasets using accelerometer data (window size 200 samples, 3 axes, 50\% overlap).}
\label{tab:motion_detection_comparison}
\end{table}

For all the techniques, we train on HHAR and finetune on RGBDGaze's training data, whose results are reported in Table~\ref{tab:motion_detection_comparison}. Our proposed SEAR-Net outperforms the other two comparison techniques, reaching an F1 score of {0.93} on HHAR and {0.80} on RGBDGaze. Compared to supervised classifiers, unsupervised detection exhibits much lower accuracy. On HHAR, VaDE-HAR achieves an F1 of 0.44, whereas Deep Clustering improves this to 0.47. On RGBDGaze, VaDE-HAR yields 0.23 F1, while Deep Clustering reaches 0.26. This is significantly below the best classification results (0.80 F1 for SEAR-Net on RGBDGaze), so we cannot rely on the unsupervised techniques for motion detection, which will generate too many false positives.

\subsubsection{Calibrator}
The continual calibration module ($M_C$) was implemented as a 2-layer MLP (256 $\rightarrow$ 128 $\rightarrow$ 2) for all experiments. We conduct grid search on the two thresholds for outlier detection $\tau$ and $\rho$ within the range $[0.75, 0.95]$ with a step size of $0.05$ and select the best setting. We have experimented with regularization-based continual learning techniques such as knowledge distillation (KD)~\cite{hinton2015distilling} along with replay, but KD does not improve the performance and adds overhead in memory and computation time. Therefore, we only opt for replay-based technique.

\section{Results}

\subsection{Overall Performance of Continual Calibration}
Our main research question is whether continual calibration improves precision of gaze prediction. To answer this question, we compare our continual calibration with the following settings:
\begin{enumerate}
\item \textit{No Calibration}: This configuration uses the output from the fine-tuned gaze estimation model ($M_V$) directly, serving as the baseline to evaluate the fundamental capability of the neural network on handheld devices. This represents the lower bound of performance in our evaluation.
\item \textit{One-off Calibration}: This involves an explicit calibration process performed once before usage, representing the state-of-the-art calibration practice. On RGBDGaze, we selected 10\% of data from each participant in a single motion condition for calibration, and then applied the adjusted model on all motions of that participant. On MotionGaze, participants performed an explicit 9-point calibration process once at the beginning, and the calibrated model was used throughout the session.
\item \textit{Oracle Motion-Aware Calibration}: This represents an ideal but impractical approach that assumes perfect knowledge of motion changes, with recalibration occurring immediately upon each motion transition. For each participant, we train a calibration model for each motion using 10\% of their respective data and use the corresponding model for each testing motion. This represents the upper bound of what can be achieved with perfect motion detection. 
\item \textit{MAC-Gaze-Classifier}: This uses SEAR-Net's classifier ($M_I$) to detect motion transitions and trigger calibration. The calibration model ($M_C$) is updated incrementally with new calibration data while preserving performance on previously seen motions through replay-based continual learning. For the memory buffer, we maintained a balanced mixture of data from all observed motions.
\end{enumerate}

\begin{table*}[!htbp]
    \centering
    \begin{subtable}{\linewidth}
    \centering
    \resizebox{0.85\textwidth}{!}{
        \begin{tabular}{lccccc}
\toprule
\textbf{Method} & \textbf{Lying} & \textbf{Sitting} & \textbf{Standing} & \textbf{Walking} & \textbf{Average} \\
\midrule
No Calibration & 1.80 (1.12) & 1.76 (1.07) & 1.70 (0.99) & 1.77 (1.07) & 1.76 (1.06) \\
One-off Calibration & 1.75 (0.96) & 1.72 (0.97) & 1.72 (0.92) & 1.74 (0.95) & 1.74 (0.96)\\
Oracle Motion-Aware & 1.53 (0.62) & 1.45 (0.65) & 1.52 (0.60) & 1.64 (0.62) & 1.54 (0.65) \\
MAC-Gaze-Classifier  & 1.55 (0.79)  & 1.56 (0.75)  & 1.59 (0.76)  & 1.63 (0.75) & 1.58 (0.79)  \\\hline
MAC-Gaze  & \textbf{1.33 (0.69)}  & \textbf{1.37 (0.63)}  & \textbf{1.43 (0.62)}  & \textbf{1.49 (0.65)} & \textbf{1.41 (0.69)}  \\
\bottomrule
\end{tabular}}
\caption{RGBDGaze}
\label{tab:overall_performance_rgbdgaze}
    \end{subtable}\par
    
\begin{subtable}{\linewidth}
    \centering
        \resizebox{0.85\textwidth}{!}{
\begin{tabular}{lcccccc}
\toprule
\textbf{Method} & \textbf{Lying} & \textbf{Sitting} & \textbf{Standing} & \textbf{Walking}& \textbf{Walking (Maze)}  & \textbf{Average} \\
\midrule
No Calibration & 2.67 (1.94) & 2.78 (1.87) & 2.76 (1.80) & 2.88 (1.88) & 2.94 (1.98) & 2.81 (2.01) \\
One-off Calibration & 2.47 (2.22)  &2.61 (2.18)  & 2.65 (2.13) & 2.82 (2.12) & 2.88 (2.30) & 2.69 (2.19) \\
Oracle Motion-Aware & 1.62 (1.51) & 2.04 (1.45) & 2.26 (1.53) & 2.53 (1.60) & 2.69 (1.73) & 2.23 (1.69) \\
MAC-Gaze-Classifier & 2.20 (1.73) & 2.33 (1.66) & 2.51 (1.67) & 2.65 (1.74) & 2.72 (1.85) & 2.48 (1.73) \\\hline
MAC-Gaze & \textbf{1.58 (1.54)} & \textbf{1.72 (1.55)} & \textbf{1.87 (1.61)} & \textbf{2.11 (1.60)} & \textbf{2.30 (1.62)} & \textbf{1.92 (1.58)} \\
\bottomrule
\end{tabular}}
\caption{MotionGaze}
\label{tab:overall_performance_ourdata}
    \end{subtable}
    \caption{Comparison of mean Euclidean errors (in cm) and standard deviations across different postures for baseline and proposed calibration strategies. Best results are highlighted in bold.}
    \label{tab:overall_performance}
\end{table*}

Table~\ref{tab:overall_performance} compares the overall performance between the above baselines and our method on RGBDGaze and MotionGaze datasets respectively. The performance is recorded in the mean and standard deviation (in brackets) of Euclidean error in cm. 
The results demonstrate that motion-aware continual calibration substantially outperforms traditional one-off calibration, particularly for motions that differ significantly from the calibration motion.

MAC-Gaze approach outperforms all the other methods. On RGBDGaze, MAC-Gaze reduces average error from 1.76\,cm (no calibration) to 1.41\,cm (average across postures), a 19.9\% improvement and on MotionGaze, it reduces error from 2.81 cm to 1.92 cm (31.7\% reduction). We observe that the performance of MAC-Gaze is more advantageous in  more dynamic conditions (walking and walking maze), where posture variations are greatest. This highlights the value of our hybrid approach in real-world mobile scenarios where users naturally shift between different postures and movement patterns. The errors of MAC-Gaze are lower than MAC-Gaze-Classifier (reducing 0.17\,cm on RGBDGaze and 0.56\,cm on MotionGaze), indicating that finer-grained motion patterns play a role in maintaining the precision of gaze tracking. 

MAC-Gaze-Classifier performs similar with Oracle Motion-Aware Calibrator on RGBDGaze as their recalibration trigger is similar; i.e., based on activity change, and RGBDGaze dataset has stable, well-distinguishable distributions between motion conditions (see in Figure~\ref{fig:datasets_acceleration_boxplots}). However, on MotionGaze, Oracle Motion-Aware Calibrator performs much better (2.23 cm vs 2.48 cm), suggesting that when the dataset exhibits a higher variety of motion distributions, the performance of gaze tracking is subject to the accuracy of the activity classifier. We will explain it more with concrete examples in the next sections. 

\textit{No Calibration} and \textit{One-off Calibration} are the worst, with One-off Calibration being marginally better. This shows the importance of continual calibration. To further understand why continual calibration is necessary, we dive deeper on One-off Calibration practice. Table~\ref{tab:oneoff_calibration_comparison} compares \emph{training motion} vs.\ \emph{testing motion} for one-off calibration. On the diagonal entries which represent scenarios where training and testing motion match, the one-off calibrator yields the lowest error of $1.54$ cm on average, confirming the importance of motion-specific calibration. In contrast, off-diagonal entries, the one-off calibrator exhibits increased errors up to 1.80 cm, highlighting significant performance degradation when calibration and usage motion differ. These results reinforce the necessity of continual calibration methods for dynamically addressing changes in head-eye-device spatial relationships encountered during realistic mobile interactions.

\begin{table}[!htbp]
\centering
\resizebox{0.7\textwidth}{!}{
\begin{tabular}{lccccc}
\toprule
 Testing & Lying & Sitting & Standing & Walking  \\
\midrule
Lying  & \textbf{1.53} (0.62) & 1.85 (1.10) & 1.83 (1.08) & 1.86 (1.09)  \\
Sitting   & 1.80 (1.12) & \textbf{1.45 }(0.65) & 1.73 (1.01) & 1.76 (1.07)\\
Standing  & 1.79 (0.99) & 1.75 (1.03) & \textbf{1.52} (0.60) & 1.71 (0.99) \\
Walking   & 1.87 (1.09) & 1.82 (1.08) & 1.77 (1.00) & \textbf{1.64} (0.62) \\
\midrule
Average & 1.75 (0.96) & 1.72 (0.97) & 1.72 (0.92) & 1.74 (0.95)\\
\bottomrule
\end{tabular}}
\caption{One-off calibration results showing the mean Euclidean error (in cm, with std in brackets) across different testing motions on RGBDGaze, when training on one motion (column) and testing on others (row).}
\label{tab:oneoff_calibration_comparison}
\end{table}

\subsection{Analysis of Calibration Trigger Frequency}

We conduct an analysis of calibration trigger frequency across both datasets. Table~\ref{tab:calibration_triggers} summarizes the number of calibration events triggered for each participant in RGBDGaze and MotionGaze datasets, alongside the data length per participant.
The results reveal significant differences in calibration patterns between the two datasets. For RGBDGaze, MAC-Gaze triggered between 4 and 9 calibrations per participant, with an average of 5.61 calibrations. This roughly matches with the expected calibration frequency, as RGBDGaze has 4 motion conditions and is expected to be recalibrated 3 times. In contrast, MotionGaze participants experienced substantially more calibration events, ranging from 10 to 31 with an average of 19.60 calibrations. This is much higher than the expected calibration frequency on MotionGaze, which has 5 motion conditions. However, the difference is because the MotionGaze dataset captures more diverse and natural movements, as shown in Figure~\ref{fig:datasets_acceleration_boxplots}. In the following, we will dive into the change detection in more details. 

\begin{table}[!htbp]
\centering
\resizebox{0.8\textwidth}{!}{
\begin{tabular}{lcccccc} 
\toprule
\textbf{Dataset} & \textbf{Data Length} & \textbf{Participants} & \textbf{Min} & \textbf{Max} & \textbf{Mean} & \textbf{Median}  \\ 
\midrule
RGBDGaze & 3,869 & 36 & 4 & 9 & 5.61 & 5 \\
MotionGaze & 17,521  & 10 & 10 & 31 & 19.60& 19\\
\bottomrule
\end{tabular}}
\caption{MAC-Gaze calibration trigger events across datasets. Data Length indicates the average number of frames per participant across all postures. Min, Max, Mean, and Median represent the distribution of calibration events triggered per participant during evaluation}
\label{tab:calibration_triggers}
\end{table}

\subsection{Change Detection}
\begin{figure}[!htbp]
    \centering
    \includegraphics[width=0.8\linewidth]{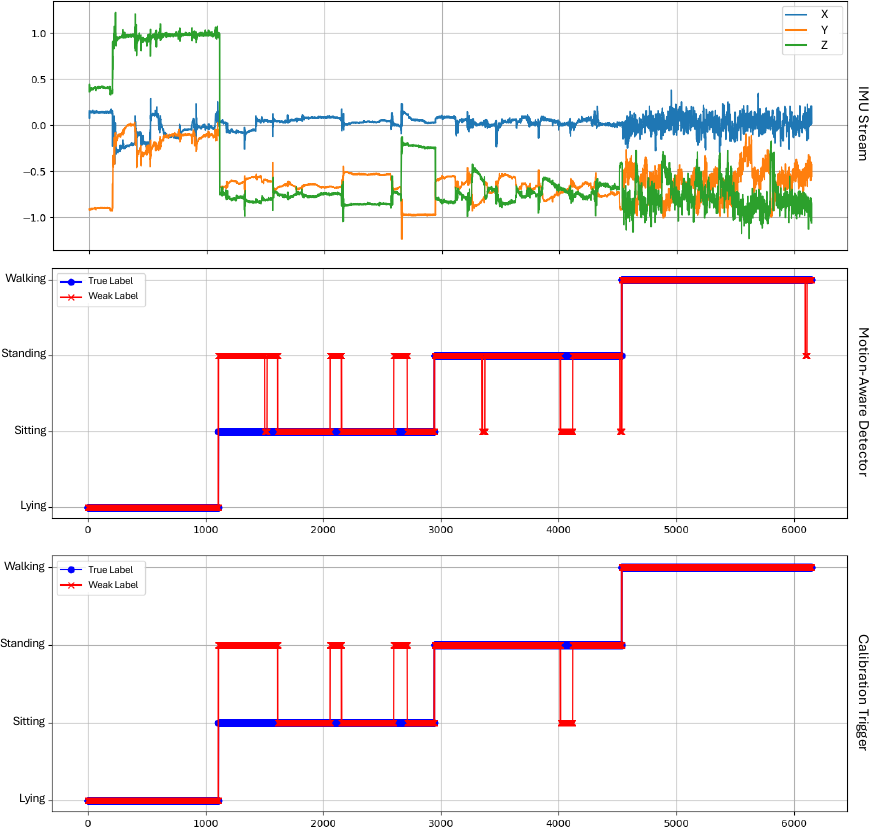}
    \caption{Temporal visualization of calibration trigger events on the RGBDGaze dataset for a participant. The upper graph shows tri-axial accelerometer data over time. The middle graph displays ground truth labels (blue) versus classifier predictions (red). The bottom graph shows calibration triggers aligned with motion detection. Blue segments indicate active motion states, while vertical red lines mark calibration events.}
    \label{fig:hte-RGBDGaze-trigger}
\end{figure}

As mentioned before, one of the key questions in continual calibration is to detect \textit{when} to trigger the calibration, with the goal to balance the number of calibrations and the performance. We examine how calibration triggers evolve over time. Figure~\ref{fig:hte-RGBDGaze-trigger} shows a representative session in RGBDGaze with transitions (e.g., from lying to walking). The \emph{4-second majority-voting window} suppresses false triggers during momentary posture shifts. We tested shorter windows (2\,s) and longer (6--8\,s); 4\,s best balanced stability with responsiveness.

\begin{figure}[!htbp]
    \centering
    \includegraphics[width=0.95\linewidth]{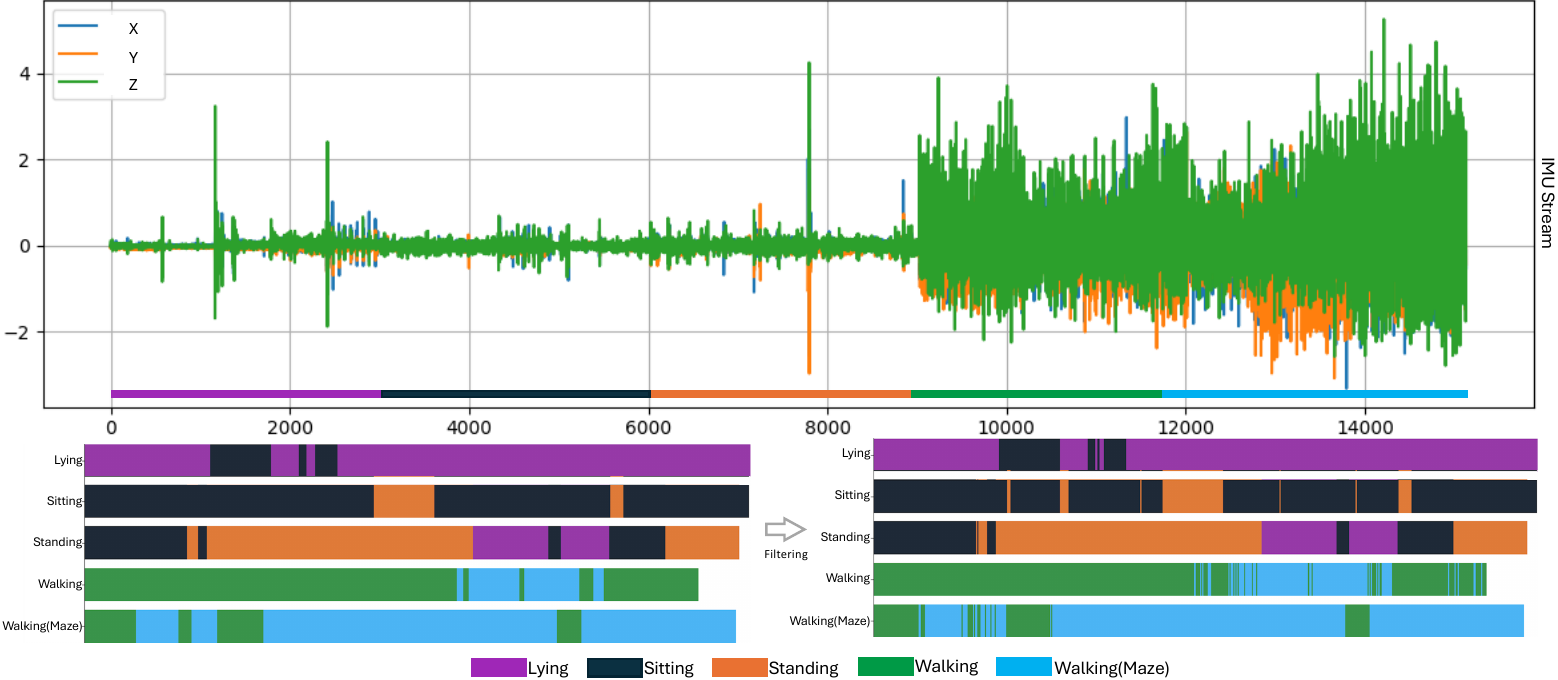}
    \caption{Temporal visualization of calibration trigger events for a participant in the MotionGaze dataset (Lying First sequence). The upper graph shows tri-axial accelerometer data displaying significantly more varied motion patterns. The middle graph shows detected postures across the sessions, with coloured bands indicating detected states and vertical red lines marking calibration events. The bottom graph displays calibration trigger timing. Compared to RGBDGaze (Figure~\ref{fig:hte-RGBDGaze-trigger}), this natural scenario exhibits more frequent and irregular motion changes.}
    \label{fig:hte-trigger-ourdata}
\end{figure}

Figure~\ref{fig:hte-trigger-ourdata}\footnote{Unlike Figure~\ref{fig:hte-RGBDGaze-trigger}, which shows ground truth vs. classifier predictions on RGBDGaze, because MotionGaze's greater length and motion diversity result in less direct alignment between HAR predictions and broad activity labels. The 4-second voting window in MAC-Gaze helps stabilize these initial HAR predictions for noisy datasets like MotionGaze.} shows these patterns for one exemplar session on MotionGaze. Compared to RGBDGaze, we observe more frequent and irregular motion change detection, reflecting more dynamic and natural interaction in real-world usage. Among motion conditions, two walking conditions are often undistinguishable, and sitting, standing, and lying are sometimes misclassified between each other. Despite this increased complexity, the 4-second majority voting mechanism still effectively prevents spurious triggers during brief postural adjustments.

The high misclassification rate helps explain why the MAC-Gaze-Classifier calibrator performs poorly on MotionGaze. If recalibration is triggered solely based on the detection of a new activity, misclassifications early in the sequence can prevent recalibration when it is actually needed. For example, in the initial tasks such as sitting, standing, and walking, the HAR model may incorrectly classify several motion conditions due to confusion among similar activities. This can lead the system to prematurely mark all motion conditions as observed. In one case, during the second task; i.e., standing, the HAR model misclassified some samples as lying, which falsely triggered a recalibration. Later, when actual lying samples were presented, the system assumed this condition had already been handled and therefore did not trigger a new recalibration. As a result, the system was calibrated on incorrect (misclassified) lying samples, leading to degraded performance on the actual lying condition.

MAC-Gaze addresses this challenge by modifying how activity change is detected. Instead of using the HAR model to identify whether a newly predicted activity has never been seen before, it simply checks whether the current prediction differs from the previous one. This reduces reliance on accurate activity classification. To further improve robustness, MAC-Gaze applies a clustering algorithm to analyze fine-grained motion patterns. Recalibration is then triggered based on whether the current motion samples deviate significantly from previously observed clusters. For instance, in the earlier example, even if the lying activity has already been marked as observed, MAC-Gaze will still trigger recalibration if GMM detects that the current lying samples differ substantially from those seen before.

\begin{table}[!htbp]
\centering
\resizebox{0.85\textwidth}{!}{%
\begin{tabular}{lc}
\toprule
\textbf{System Configuration} & \textbf{Mean Error (cm)} \\
\midrule
Full MAC-Gaze System ($M_V + M_I + \text{Replay} + \text{Hybrid Trigger}$) & \textbf{1.41 (0.69)} \\
Without Hybrid Trigger ($M_V + M_I + \text{Replay} + M_I \text{ Classifier Only}$) & 1.58 (0.79) \\
Without Replay ($M_V + M_I + \text{Hybrid Trigger}$) & 1.73 (1.19) \\
Without Motion-based Trigger ($M_V + \text{Time-based Replay every 30s}$) & 1.71 (1.46) \\
\bottomrule
\end{tabular}}
\caption{Ablation study results on the RGBDGaze dataset, showing the impact of removing key components from the full MAC-Gaze system. Error values represent mean Euclidean error (cm) with standard deviation in parentheses, averaged across all postures. The "Hybrid Trigger" combines $M_I$'s classifier output with GMM clustering; "Without Hybrid Trigger" relies solely on $M_I$'s classifier.}
\label{tab:ablation_components}
\end{table}

To understand the contribution of each component, we conducted ablation studies by selectively replacing or removing components of our system. Table~\ref{tab:ablation_components} quantifies the impact of different system configurations on overall performance.  The \textit{Without Replay} condition removes the memory buffer for continual learning, showing a significant performance drop; that is, the error increases from 1.41 cm to 1.73 cm. This demonstrates the importance of replay-based learning in mitigating catastrophic forgetting. The \textit{Without Motion Detection} condition applies calibration at fixed time intervals rather than based on detected motion changes. To ensure a fair comparison, we implemented a heuristic approach where each participant received the same number of calibrations as they would have with MAC-Gaze, but distributed at regular intervals throughout their session. For example, if a participant experienced 5 recalibrations with MAC-Gaze during a 15-minute session, the \textit{Without Motion Detection} condition would trigger recalibrations every 3 minutes regardless of the participant's actual movements or posture changes. Despite having the same number of calibration, the time-based approach showed the worst performance (1.71 cm vs. 1.41 cm), which  highlights that \textit{when} calibration occurs is just as important as how frequently it occurs.

\subsection{Replay Ratio Impact}
We further test the impact of different replay ratios on RGBDGaze in Table~\ref{tab:rgbdgaze_perf_reply}. The replay ratio of 70\% yields the lowest average error (1.59 cm) across all initial postures, with particularly strong performance (1.56 cm) when starting from standing or walking. We observe that very low replay ratios (10\%) provide insufficient memory of previous tasks, leading to catastrophic forgetting, particularly evident in Lying First and Standing First conditions. High replay ratios (80-90\%) over-represent previous tasks at the expense of adapting to new postures, resulting in degraded overall performance.

Our analysis also reveals consistent performance across different task sequence orderings. Examining the initialization conditions across columns in Table~\ref{tab:rgbdgaze_perf_reply}, we observe similar mean errors across all starting conditions: Lying First ($M=1.63$ cm, $SD=0.068$ cm), Sitting First ($M=1.65$ cm, $SD=0.058$ cm), Standing First ($M=1.66$ cm, $SD=0.076$ cm), and Walking First ($M=1.64$ cm, $SD=0.061$ cm). One-way ANOVA confirms no significant difference between these sequence orders ($F=0.36$, $p=0.782$), indicating that MAC-Gaze maintains consistent performance regardless of the initial posture. This robustness to starting conditions suggests that our approach effectively adapts to various motion patterns, regardless of their order of occurrence.

\begin{table}[!htbp]
\centering
\resizebox{0.8\textwidth}{!}{
\begin{tabular}{cccccc}
\toprule
{Replay Ratio} & Lying First & Sitting First & Standing First & Walking First  & Average  \\\hline
10\%                                    & \cellcolor[HTML]{FDBB7B}1.69          & \cellcolor[HTML]{AAD27F}1.59          & \cellcolor[HTML]{FECA7E}1.67          & \cellcolor[HTML]{FCB079}1.70          & 1.66                                   \\
20\%                                    & \cellcolor[HTML]{C4DA80}1.60          & \cellcolor[HTML]{FCA978}1.71          & \cellcolor[HTML]{FDBA7B}1.69          & \cellcolor[HTML]{A7D17E}1.59          & 1.65                                  \\
30\%                                    & \cellcolor[HTML]{A3D07E}1.59          & \cellcolor[HTML]{FB9C75}1.73          & \cellcolor[HTML]{87C87D}1.57          & \cellcolor[HTML]{FBA076}1.72          & 1.65                                \\
40\%                                    & \cellcolor[HTML]{FFE984}1.64          & \cellcolor[HTML]{A2D07E}1.59 & \cellcolor[HTML]{FFE082}1.65          & \cellcolor[HTML]{FDC27D}1.68          & 1.64                         \\
50\%                                    & \cellcolor[HTML]{CDDC81}1.61          & \cellcolor[HTML]{FFE683}1.64          & \cellcolor[HTML]{D8DF81}1.61          & \cellcolor[HTML]{FFE583}1.64          & 1.62                       \\
60\%                                    & \cellcolor[HTML]{6DC17B}1.56 & \cellcolor[HTML]{FCEA83}1.63          & \cellcolor[HTML]{FDEA83}1.63          & \cellcolor[HTML]{FDB67A}1.70          & 1.63                      \\
\textbf{70\%}                           & \cellcolor[HTML]{AFD47F}1.59          & \cellcolor[HTML]{FFEB84}1.63          & \cellcolor[HTML]{6CC07B}1.56 & \cellcolor[HTML]{63BE7B}1.56 & \textbf{1.59}              \\
80\%                                    & \cellcolor[HTML]{A9D27F}1.59          & \cellcolor[HTML]{E0E282}1.62          & \cellcolor[HTML]{FB9373}1.74          & \cellcolor[HTML]{B1D47F}1.59          & 1.64                 \\
90\%                                    & \cellcolor[HTML]{F96F6D}1.78          & \cellcolor[HTML]{FB9273}1.74          & \cellcolor[HTML]{F8696B}1.79          & \cellcolor[HTML]{8DCA7D}1.58          & 1.72                                \\ 
\midrule
Average & 1.63 (0.068) & 1.65 (0.058) & 1.66 (0.076) & 1.64 (0.061)  & -  \\
\bottomrule
\end{tabular}}
\caption{Effect of different replay ratios on mean  Euclidean error (cm) for MAC-Gaze on RGBDGaze, showing errors for different initial calibration postures and their sequences. Colour gradient from green (lower error) to red (higher error) visualizes performance differences.}
\label{tab:rgbdgaze_perf_reply}
\end{table}

\subsection{Generalisation to Different Base Gaze Models}
Our technique allows plug-and-play integration with different base gaze estimation models. Table~\ref{tab:overall_performance_ourdata_itracker} presents results with iTracker~\cite{krafka2016eye}, a well-established mobile gaze estimation model. The results demonstrate that MAC-Gaze maintains its effectiveness when applied to iTracker, despite its generally higher baseline error rates. On RGBDGaze, MAC-Gaze reduces iTracker's error from 2.77\,cm to 2.16\,cm (22.0\% improvement), comparable to the 19.9\% reduction with MViTGaze. On MotionGaze, MAC-Gaze achieves a 40.1\% reduction (4.49\,cm to 2.69\,cm).

MAC-Gaze with iTracker outperforms the Oracle Motion-Aware approach (2.16\,cm vs. 2.48\,cm) on RGBDGaze and it shows the largest improvements in dynamic conditions; for example, on RGBDGaze, the error for walking is reduced from 2.79\,cm to 2.27\,cm, leading to 18.6\% improvement. The consistent performance improvements across both base models highlight the generalizability of our continual calibration approach, suggesting our method can enhance a wide range of existing gaze estimation systems without requiring fundamental architectural changes. Furthermore, MAC-Gaze outperforms MAC-Gaze-Classifier by a larger margin when applied to iTracker (16.6\% improvement on RGBDGaze) compared to MViTGaze (10.8\% improvement), suggesting that fine-grained motion pattern detection becomes increasingly valuable for models with higher baseline error. This may be because iTracker's representations are less robust to posture variations, making it particularly beneficial to precisely identify when recalibration is needed and adapt to subtle changes in head-eye-camera relationships.

\begin{table*}[ht]
\centering
\begin{subtable}{\linewidth}
\centering
\resizebox{0.85\textwidth}{!}{
\begin{tabular}{lccccc}
\toprule
\textbf{Method} & \textbf{Lying} & \textbf{Sitting} & \textbf{Standing} & \textbf{Walking} & \textbf{Average} \\
\midrule
No Calibration (iTracker) & 2.85 (2.77) & 2.66 (2.76) & 2.78 (2.74) & 2.79 (2.82) & 2.77 (2.81) \\
One-Off Calibration (iTracker) & 2.83 (2.53) & 2.61 (2.59) & 2.70 (2.67) &  2.84 (2.66) &  2.74 (2.64) \\
Oracle Motion-Aware (iTracker) & 2.42 (1.93) & 2.37 (1.95) & 2.52 (2.13) & 2.62 (2.13) & 2.48 (1.98) \\
MAC-Gaze-Classifier (iTracker) & 2.44 (2.13) & 2.46 (2.18) & 2.59 (2.27) & 2.72 (2.25)  & 2.59 (2.24) \\
\hline
MAC-Gaze (iTracker) & \textbf{2.03 (1.33)} & \textbf{2.16 (1.38)} & \textbf{2.19 (1.37)} & \textbf{2.27 (1.45)}  & \textbf{2.16 (1.44)} \\
\bottomrule
\end{tabular}}
\caption{RGBDGaze Dataset Performance (with iTracker)}
\label{tab:overall_performance_itracker_rgbdgaze}
\end{subtable}\par

\begin{subtable}{\linewidth}
\centering
\resizebox{0.85\textwidth}{!}{
\begin{tabular}{lcccccc}
\toprule
\textbf{Method} & \textbf{Lying} & \textbf{Sitting} & \textbf{Standing} & \textbf{Walking}& \textbf{Walking (Maze)}  & \textbf{Average} \\
\midrule
No Calibration (iTracker) & 4.12 (2.99) & 4.41 (2.77) & 4.48 (2.94) & 4.54 (3.06) & 4.88 (3.23) & 4.49 (3.27) \\
One-Off Calibration (iTracker) & 3.53 (2.18) & 3.08 (1.98) & 3.25 (2.09) & 3.56 (2.30) & 3.79 (2.44) & 3.45 (2.19) \\
Oracle Motion-Aware (iTracker) & 2.38 (2.12) & 2.31 (1.99) & 2.77 (2.09) & 3.18 (2.20) & 3.37 (2.45) & 2.86 (2.42) \\
MAC-Gaze-Classifier (iTracker) & 2.49 (2.01) & 2.46 (2.05) & 2.78 (2.35) & 3.26 (2.46) & 3.44 (2.19) & 2.90 (2.15) \\
\hline
MAC-Gaze (iTracker) & \textbf{2.31 (1.97)} & \textbf{2.22 (1.99)} & \textbf{2.59 (2.11)} & \textbf{3.00 (2.14)} & \textbf{3.32 (2.05)} & \textbf{2.69 (2.06)} \\
\bottomrule
\end{tabular}}
\caption{MotionGaze Dataset Performance (with iTracker)}
\label{tab:overall_performance_itracker_motiongaze}
\end{subtable}
\caption{Mean Euclidean error (cm) using iTracker as the base model, with standard deviations in parentheses, across different postures for baseline and proposed calibration strategies. Best results are highlighted in bold.}
\label{tab:overall_performance_ourdata_itracker}
\end{table*}

\section{Discussion}

\subsection{Key Findings and Implications}
Our experimental results confirm that MAC-Gaze effectively address the calibration distortion problem in mobile gaze tracking and provides highly accurate gaze tracking on handheld mobile devices. MAC-Gaze outperforms the Oracle Motion-Aware approach on RGBDGaze (1.41 cm vs. 1.54 cm) and the fixed time interval calibration on RGBDGaze (1.41 cm vs. 1.71 cm), despite the Oracle method having priori knowledge of posture transitions and the fixed time calibration has multiple recalibration times. This suggests that our fine-grained motion pattern detection captures more nuanced variations in head-eye-camera and the motion-caused input uncertainty during user-device interaction than discrete activity classes alone.

The ablation studies validate each component's importance in our system. Removing the hybrid clustering component increases error by 17.4\%, while eliminating replay-based continual learning degrades performance by 22.7\%. This indicates that both accurate motion detection and effective knowledge retention are critical to the system's success.

The one-off calibration results (Table~\ref{tab:oneoff_calibration_comparison}) empirically confirm our hypothesis that changes in the spatial relationship between head, eyes, and camera significantly impact calibrated gaze estimation accuracy. When calibration and current motion does not match, errors increase by 17.1\% on average (1.54 cm to 1.80 cm), which provides the empirical evidence on detecting motion changes is crucial for accuracy and important for timely recalibration.

\subsection{Effectiveness of Change Detection}
The superior performance of our hybrid approach over purely supervised methods stems from its ability to detect both coarse-grained activity transitions and fine-grained variations within activities. The comparison of clustering approaches in Figure~\ref{fig:clustering_comparison} reveals that raw IMU signals provide better discrimination of motion patterns than latent features from the encoder. While the encoder excels at activity classification, it may compress subtle variations that impact gaze estimation. By operating directly on raw sensor readings, our clustering approach preserves these fine-grained distinctions.

The temporal analysis of calibration triggers (Figures~\ref{fig:hte-RGBDGaze-trigger} and \ref{fig:hte-trigger-ourdata}) demonstrates the effectiveness of our 4-second majority voting mechanism in suppressing spurious triggers while maintaining responsiveness to genuine transitions. The decreasing trigger frequency over time indicates the system's ability to accumulate knowledge about diverse motion conditions, reducing the need for frequent recalibration.

\begin{figure}[!htbp]
    \centering
    \includegraphics[width=0.7\linewidth]{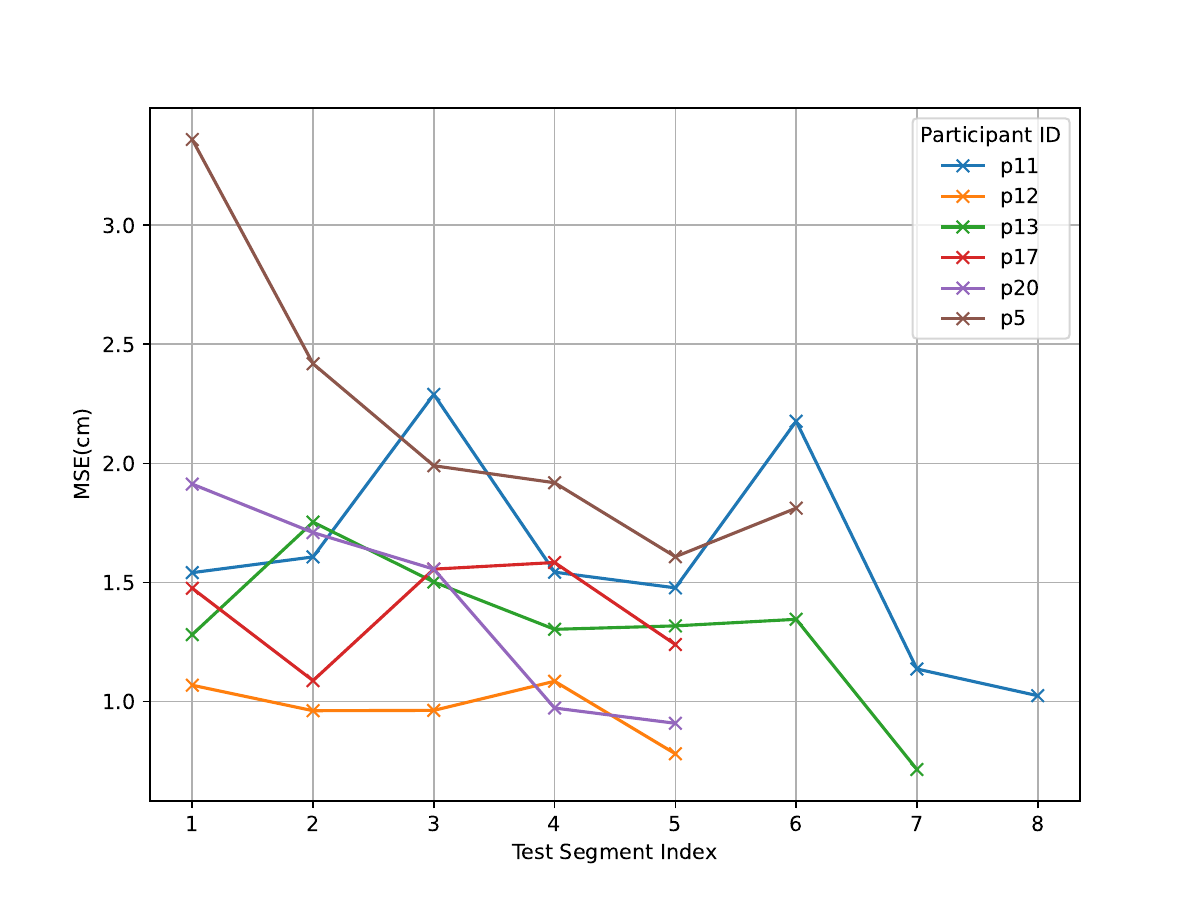}
    \caption{Temporal visualization of gaze estimation performance across test segments for multiple participants on the RGBDGaze dataset using MAC-Gaze. Each test segment represents the sequential order of recalibration events.}
    \label{fig:MAC-Gaze-acc-trend}
\end{figure}

Figure~\ref{fig:MAC-Gaze-acc-trend} illustrates the progressive performance of MAC-Gaze across multiple recalibration cycles for several participants from the RGBDGaze dataset. Each point on the graph represents the mean squared error (MSE) measured during a test segment—the period following a calibration event where the newly updated model is applied for gaze estimation before the next recalibration occurs. For most participants (e.g., p5, p12, and p20), we observe a general trend of decreasing error as the system accumulates knowledge through successive recalibrations, demonstrating the effectiveness of our continual learning approach. However, participants p11 and p17 exhibit notable fluctuations in error rates, with occasional increases following recalibration. Upon analysing their motion data, we discovered that these participants demonstrated more frequent and abrupt changes in device handling compared to others. This creates a temporal mismatch between motion detection and model updating; by the time the system completes recalibration, their motion state has already transitioned to a new pattern not represented in the calibration data. This observation highlights a challenge in real-world deployments where users with highly dynamic interaction patterns may require more responsive calibration mechanisms or predictive motion modelling.

\subsection{Limitations and Future Directions}
Despite its effectiveness, MAC-Gaze faces several practical challenges. First of all, we have observed high calibration frequency on MotionGaze in Table~\ref{tab:calibration_triggers}, which can cause disruption to user experience. There are several reasons behind the high frequency. First of all, when we collect the data, we encourage participants to change their device holding posture naturally. We hope the short-term data can represent the long-term data in real-world scenarios. That's why our MotionGaze data incurs diverse motion distribution than RGBDGaze, and leads to much higher calibration frequency. Secondly, our motion detection algorithm needs improvement. Constrained by the number of calibration and replay samples, we have tried temporal voting to suppress the spurious motion detection and GMM to detect finer-grained motion change, which works to a certain degree. In the future, we need to look into more advanced techniques to improve HAR accuracy; for example, mixup~\cite{lu2022semantic} or unsupervised domain adaptation~\cite{hu2023swl} to enable more accurate adaptation to the new dataset. Thirdly, the HAR model trained on HHAR dataset which only have 6 activities, which maybe limit the pattern's recognition. In the following research, the activity classifier could be plug-and-play adapted various architectures and trained on the merged dataset based on various open-source datasets to make the activity recognition more fine-course.

The computational overhead of continuous motion monitoring may impact battery life on mobile devices, necessitating more efficient implementations. Our reliance on explicit calibration when motion changes are detected interrupts the user experience, suggesting integration with implicit calibration, such as mouse click events on desktop~\cite{sugano2015self}, and touch events in mobile device~\cite{cai2025gazeswipe} methods as a valuable future direction.

\section{Conclusion}
This paper has presented MAC-Gaze, a novel motion-aware continual calibration approach for mobile gaze tracking that addresses the fundamental challenge of maintaining gaze estimation accuracy amid dynamic user-device interactions. By leveraging IMU sensors to detect changes in motion states and employing continual learning techniques to adapt to evolving conditions, our approach significantly improves gaze tracking performance across static and dynamic scenarios. The experimental results demonstrate substantial improvements over traditional one-off calibration methods, with error reductions of 19.9\% on RGBDGaze and 31.7\% on MotionGaze datasets. Our ablation studies further validate the importance of each component, with the hybrid decision-making approach and replay-based continual learning proving essential to the system's success. The demonstrated generalizability of MAC-Gaze across different backbone models confirms its potential as a universal enhancement for appearance-based gaze estimation systems, with this adaptability being particularly valuable for deployment across various device configurations and application contexts. Our analysis of replay ratios provides practical guidance for implementing similar systems, with the 70\% ratio offering optimal balance between retaining previous knowledge and adapting to new conditions. Despite these advances, several challenges remain for future work: the computational overhead of continuous motion monitoring necessitates more efficient implementations for resource-constrained mobile devices; the current reliance on explicit calibration interrupts the user experience, suggesting integration with implicit calibration methods as a valuable direction; and exploring more sophisticated continual learning techniques with smaller memory footprints, incorporating additional sensing modalities, and developing personalization approaches that adapt to individual users' behaviour patterns over time could further enhance performance and usability. In conclusion, MAC-Gaze represents a step towards robust, adaptive gaze tracking on mobile devices, addressing a critical obstacle to wider adoption of this technology in everyday interactions and bringing the potential of gaze-based interfaces closer to practical reality.

\section*{Acknowledgements}
We thank Dr. Yafei Wang and Dr. Kaixing Zhao for their valuable feedback on the paper revision.

\bibliographystyle{acm}
\bibliography{references}  %%% Uncomment this line and comment out the ``thebibliography'' section below to use the external .bib file (using bibtex) .





% \end{thebibliography}

\end{document}